\documentclass[10pt,onecolumn,showpacs,aps,prd,preprintnumbers,amsmath,amssymb,superscriptaddress,nofootinbib,postscript]{revtex4-1}
\usepackage{enumerate}
\usepackage{epsfig} 
\usepackage{float} 
\usepackage[caption = false]{subfig}
\usepackage{wasysym} 
\usepackage{mathrsfs} 
\usepackage{amsfonts} 
\usepackage{amsbsy} 
\usepackage{amscd} 
\usepackage{pstricks} 
\usepackage{multirow} 
\usepackage{tikz}
\usepackage{color}
\usepackage{slashed}
\usepackage{array}
\usepackage{tablefootnote} 
\usepackage{enumitem}
\usepackage[utf8]{inputenc}

\usepackage{color}
\definecolor{myred}{rgb}{0.6,0,0} 
\definecolor{myblue}{rgb}{0,0.2,0.4}
\definecolor{mygreen}{rgb}{0,0.9,0.1}
\definecolor{hc}{rgb}{.9,0.1,0.7}
\definecolor{hcout}{rgb}{.9,0.7,0.9}
\definecolor{Orange}{rgb}{1.,0.65,0.}

\makeatletter

\newcommand{\fmslash}[2][0mu]{%
  \mathchoice
    {\fmsl@sh\displaystyle{#1}{#2}}%
    {\fmsl@sh\textstyle{#1}{#2}}%
    {\fmsl@sh\scriptstyle{#1}{#2}}%
    {\fmsl@sh\scriptscriptstyle{#1}{#2}}}
\newcommand{\fmsl@sh}[3]{%
  \m@th\ooalign{$\hfil#1\mkern#2/\hfil$\crcr$#1#3$}}
\makeatother

\newcommand{\lsim}{{\;\raise0.3ex\hbox{$<$\kern-0.75em\raise-1.1ex\hbox{$\sim$}}\;}}
\newcommand{\gsim}{{\;\raise0.3ex\hbox{$>$\kern-0.75em\raise-1.1ex\hbox{$\sim$}}\;}}

\usepackage{array}
\newcolumntype{C}[1]{>{\centering\arraybackslash$}p{#1}<{$}}

\usepackage{bm} 
\newcommand{\be}{\begin{equation}}
\newcommand{\ee}{\end{equation}}
\newcommand{\bes}{\begin{equation*}}
\newcommand{\ees}{\end{equation*}}
\newcommand{\bea}{\begin{eqnarray}}
\newcommand{\eea}{\end{eqnarray}}
\newcommand{\beas}{\begin{eqnarray*}}
\newcommand{\eeas}{\end{eqnarray*}}

\newcommand{\snu}{\tilde{\nu}}

\begin{document}
\title{Right sneutrino with $\Delta\,L\,=\,2$ masses as non-thermal dark matter} 

\author{Avirup Ghosh}
\email{avirupghosh@hri.res.in}
\affiliation{Regional Centre for Accelerator-based Particle Physics,
Harish-Chandra Research Institute, HBNI,
Chhatnag Road, Jhunsi, Allahabad - 211 019, India} 
\author{Tanmoy Mondal} 
\email{tanmoymondal@hri.res.in}
\email{tanmoy@kias.re.kr}
\affiliation{Regional Centre for Accelerator-based Particle Physics,
Harish-Chandra Research Institute, HBNI,
Chhatnag Road, Jhunsi, Allahabad - 211 019, India} 
\affiliation{School of Physics, Korea Institute for Advanced Study, Seoul 130-722, Republic of Korea}
\author{Biswarup Mukhopadhyaya}
\email{biswarup@hri.res.in} 
\affiliation{Regional Centre for Accelerator-based Particle Physics,
Harish-Chandra Research Institute, HBNI,
Chhatnag Road, Jhunsi, Allahabad - 211 019, India} 

\date{\today}
\preprint{HRI-RECAPP-2018-005}
\preprint{KIAS-P19011}

\begin{abstract}
We consider MSSM with right-chiral neutrino superfields with Majorana masses, where the 
lightest right-handed sneutrino dominated scalars constitutes non-thermal dark matter (DM). 
The $\Delta\,L=2$ masses are subject to severe constraints coming from freeze-in 
relic density of such DM candidates as well as from sterile neutrino 
$\textit{freeze-in}$.
In addition, big-bang Nucleosynthesis and $\textit{freeze-out}$ of the next-to-lightest superparticle 
shrink the viable parameter space of such a scenario. We examine various $\Delta\,L=2$ mass 
terms for families other than that corresponding to the LSP sneutrino.
$\Delta\,L=2$ masses are difficult to reconcile with a right-sneutrino DM, unless there is either 
(a) a hierarchy of about 3 orders of magnitudes among various supersymmetry-breaking mass parameters,
or, (b) strong cancellation between the higgsino mass and the trilinear supersymmetry breaking 
mass parameter for sneutrinos.
\end{abstract}


\pacs{14.80.Ly,12.60.Jv, 95.60.Cq} 
\maketitle


\section{Introduction} \label{sec:intro}
While the search for the dark matter (DM) candidate(s) of our universe is on,
one constantly feels the necessity of going beyond stereotypes in modelling
physics beyond the standard model (SM) to accommodate the candidate 
particle(s). It is in this spirit that alternative candidates in small 
extensions of the minimal supersymmetric standard model (MSSM) have been 
explored. One example consists in scenarios with gravitino as warm DM~\cite{Arcadi:2015ffa, Arvey:2015nra, Catena:2014pca, Covi:2014roa, Covi:2012wx, Covi:2009bk, Dutta:2017jpe, Ferrantelli:2017ywq, Munoz:2017ezd}.
Another, quite minimalistic, extension is to extend the MSSM with a right-chiral neutrino superfield for each 
family, and postulate that one right-sneutrino-dominated scalar is the DM 
candidate. Various cosmological as well as phenomenological implications of 
this scenario have already been explored
~\cite{Munoz:2017ezd, Banerjee:2018uut, DelleRose:2018rpp, Chatterjee:2017nyx, Banerjee:2016uyt, Arina:2015uea, CabreraCatalan:2015jjy, Abdallah:2018gjj, Banerjee:2013fga}. 

The interaction of a sneutrino DM particle with all other MSSM fields
is proportional to the very small neutrino Yukawa coupling.
Consequently, this scenario almost always leads to a
non-thermal DM, with a long-lived next-to-lightest SUSY particle
(NLSP). Such a long-lived NLSP mostly survives till it decouples from
the thermal bath, and decays to the DM candidate thereafter. Such
scenarios have been often explored {\it by assuming that neutrinos 
have just Dirac masses}, in which cases their Yukawa coupling strengths are 
${\cal{O}}(10^{-13})$~\cite{Biswas:2009rba, Biswas:2009zp, Biswas:2010cd, Choudhury:2008gb, Gupta:2007ui, Banerjee:2016uyt, Asaka:2005cn, Asaka:2006fs, Asaka:2007zza}.

Some additional issues become important if neutrinos have Majorana
masses as well. This will require the right-handed neutrino fields to
have $\Delta\,L\,=\,2$ mass terms. The Type-I seesaw mechanism works
here, thus requiring somewhat higher neutrino Yukawa couplings. This,
however, entails the process of freeze-in of right sneutrinos while
the NLSP (a stau, for example) is yet to decouple. Studies including
those on similar non-SUSY theories have shown that such freeze-in
contribution to the relic density impose rather strong constraints on
the corresponding model parameters~\cite{Molinaro:2014lfa, Hessler:2016kwm,Ghosh:2017vhe, Borah:2017dfn}.
In MSSM with right sneutrino DM and Dirac masses, too, this effect is 
small but non-negligable and it constrains the associated parameter 
space~\cite{ Asaka:2005cn, Asaka:2006fs, Asaka:2007zza,Banerjee:2018uut}.
With $\Delta L = 2$ mass terms present, however, the seesaw formula allows 
bigger Yukawa couplings, and thus the constraints tighten. 
For a degenerate light neutrino spectrum around 0.1 eV the limit on 
Majorana masses and Yukawa coupling was studied in~\cite{Gopalakrishna:2006kr}.
These observations were restricted to specific scenarios where
all relevant SUSY-breaking parameters were taken to be around the 
electroweak scale ({\it i.e.} 100 GeV). In \cite{Yaguna:2008mi} it was shown that 
the near degeneracy among the NLSP and LSP decreases the relic density due 
to inverse decay of LSP. 

The present study is more general, and takes into account wider ranges 
of mass scales as well the interplay of various SUSY-breaking parameters.
Unlike most approaches {\it we allow a hierarchy of SUSY-breaking mass 
parameters}. The freeze-in of right-sneutrino DM from the decay of heavier 
superparticles will be affected by the (small but non-vanishing) 
left-sneutrino component in the DM and hence the
deciding factor is the sneutrino mixing angle, which essentially depends
on the off-diagonal part of the sneutrino mass matrix and is tightly 
constrained. One has to also fit the neutrino mass and mixing patterns~\cite{Esteban:2016qun}, and, quite seriously, the potential
contribution of long-lived sterile neutrinos generated via the
Dodelson-Widrow mechanism~\cite{Dodelson:1993je, Shakya:2015xnx, Boyarsky:2008ju, Gorbunov:2008ka, Essig:2013goa, Roy:2010xq, Canetti:2010aw, Asaka:2006nq}. 
Thus the $\Delta L = 2$ masses, {\it{especially those corresponding to 
the families other than the one the DM belongs to}}, get seriously 
restricted, especially in view of the present limit on the dark matter relic 
density by the PLANCK collaboration~\cite{Ade:2015xua}. The way such 
restrictions arise is  investigated in the present paper.

There can be several scales associated with the $\Delta L = 2$ masses,
consistent with the Type-I seesaw mechanism. The commonly known GUT-scale 
seesaw won't work for a case where non-thermal DM candidates are sought, as 
that would entail Yukawa couplings large enough for them to thermalise. 
We find that, Majorana masses in the electroweak scale, too, allow
Yukawa couplings consistent with freeze-in rate, if a suppression in 
left-right mixing occurs via  diagonal SUSY breaking terms that are several 
orders of magnitude larger than the off-diagonal ones. Though the situation 
is marginally better for ${\cal{O}}(1)$ GeV Majorana masses, a fine-tuned 
cancellation between the soft-and F-terms is required there as well. 
The scenario with all the three Majorana masses ranging from 500 MeV
down to a few tens of keV is constrained by light element abundance. 
When all three masses are in keV scale, they all become warm
dark matter, a scenario ruled out already due to overproduction of the
warm keV scale dark matter via Dodelson-Widrow mechanism. We may have one 
$\Delta\,L=2$ mass in the eV scale~\cite{deGouvea:2005er}, as 
considered in~\cite{Gopalakrishna:2006kr}, but such a situation brings in 
severe cosmological constraints~\cite{Cyburt:2015mya, Ade:2015xua}.
We establish $\textit{freeze-in}$ of right-sneutrino DM to be a major
constraining factor with $\Delta\,L=2$ masses. In such a case, not only the 
NLSP but also the rest of the MSSM spectrum contributes to 
$\textit{freeze-in}$. This constraint is thus largely applicable to scenarios 
including various NLSP candidates. We focus on the stau and neutralino
NLSP in particular. 

Our paper have been organized as follows. In section~\ref{sec:model} 
we discuss the model considered by us and the related constrains coming from 
$\textit{freeze-in}$, NLSP $\textit{freeze-out}$, low mass sterile neutrinos 
and Big-bang Nucleosynthesis (BBN). Section~\ref{sec:strategy} has been 
devoted to the discussions of the analysis strategy followed by us in 
determining the allowed parameter space still allowed from the constraints
mentioned in section~\ref{sec:model}. 
In section~\ref{sec:details1} we have shown how the constraints mentioned 
already can severely restrict the parameter space of such a scenario following 
the strategy mentioned. We summarise and conclude in 
section~\ref{sec:summary}. Finally the formula used by us have been tabulated 
in Appendices \ref{sec:Appendix1} and ~\ref{sec:Appendix2}. 

\section{The Model and constraints}\label{sec:model}
We start by outlining the salient features of a right sneutrino DM 
scenario with $\Delta\,L=2$ masses, which warrant a fresh study of the 
constraints at focus.
We consider the MSSM scenario augmented with three right chiral neutrino
superfields ($\hat{N}_{R}$), which possess $\Delta\,L=2$ mass terms.
Hence the superpotential gets extended to the form~\cite{Hirsch:1997vz, Hall:1997ah, Grossman:1997is}
\begin{equation}\label{eqn:superpotential}
\mathcal{W} = \mathcal{W}_{MSSM} + Y_{\nu}^{ij} \hat{H}_{u}\hat{L}^{i}\hat{N}
_{R}^{j} + \frac{1}{2}M_{N}^{ij}\hat{N}_{R}^{i}\hat{N}_{R}^{j},
\end{equation}
where $\hat{H}_{u}$ is the higgs doublet that couples to up-type quarks,
$\hat{L}$ is left-chiral SU(2) doublet lepton superfields and
$\hat{N}_{R}$ is a right-handed neutrino superfield. While $Y_{\nu}$ is 
the neutrino Yukawa coupling matrix, $M_{N}$ is the $\Delta\,L\,=\,2$ mass 
matrix for the heavy neutrinos, assumed, as already stated, to be diagonal, 
since basis rotations in the right-handed neutrino sector is unlikely to 
affect our main conclusions.

Since we are not assuming any high-scale mechanism of SUSY breaking 
we need to add all allowed SUSY-breaking terms phenomenologically.
The relevant soft terms in this case are
\begin{equation}\label{eqn:soft-terms}
\mathcal{L}_{soft} = \mathcal{L}_{soft}^{MSSM} - m_{N}^{2\,ij}\tilde{N}_{R}^{*i}
\tilde{N}_{R}^{j}+ (m_{B}^{2\,ij}\tilde{N}_{R}^{i}\tilde{N}_{R}^{j} - T_{\nu}
^{ij}h_{u}\tilde{L}^{i}\tilde{N}_{R}^{j} + h.c.),
\end{equation}
where $m_{N}$($m_{B}$) corresponds to $\Delta\,L=0(2)$ susy-breaking 
sneutrino masses and $T_{\nu}$ is the coefficient of the trilinear 
SUSY-breaking term. The left-right mixing in the sneutrino sector occurs, via 
the soft terms proportional to $T_{\nu}$, and the F-terms proportional to 
the higgsino mass parameter($\mu$), when $h_{u}(h_{d})$ acquires the 
vacuum expectation values $v_{u}(v_{d})$. In addition to $m_{N}$ and 
$m_{B}$, the F-term masses proportional to $M_{N}$ adds to the 
right-handed sneutrino masses. In our case, one requires $M_{N}$ not to 
exceed this SUSY-breaking scale, so that a right sneutrino may behave as a 
DM candidate.

To study the sneutrino mass terms and phenomenology of sneutrinos it is convenient to introduce the real fields, ($\snu^{i}_{1}\,,\,\tilde{N}^{i}_{1}\,,\,\snu^{i}_{2}\,,\,\tilde{N}^{i}_{2}$) defined as follows~\cite{Yaguna:2008mi, Gopalakrishna:2006kr, Grossman:1997is}:
\begin{equation}
\snu^{i}_{L}\,=\,\frac{1}{\sqrt{2}}(\snu^{i}_{1}\,+\,i\,\snu^{i}_{2}),\,
\tilde{N}^{i}_{R}\,=\,\frac{1}{\sqrt{2}}(\tilde{N}^{i}_{1}\,+\,i\,\tilde{N}^{i}_{2}),
\label{realsnu1}
\end{equation}

In the basis constituted by CP-even($\snu_{1}\,,\,\tilde{N}_{1}$) 
and CP-odd ($\snu_{2}\,,\,\tilde{N}_{2}$) real sneutrino fields, 
assuming no CP-violation in the SUSY sector the 
sneutrino mass matrix takes the block-diagonal form
\begin{equation}
M_{\tilde{\nu}} =
\begin{bmatrix}
 m_{LL}^{2} & m_{RL}^{2} + m_{D}M_{N} &  0 & 0  \\
 m_{RL}^{2\,T}+ M_{N}^{\,T}m_{D}^{\,T} & m_{RR}^{2}-m_{B}
 ^{2} & 0 & 0 \\
 0 & 0 & m_{LL}^{2} & m_{RL}^{2}- m_{D}M_{N} \\
 0 & 0 & m_{RL}^{2\,T}- M_{N}^{\,T}m_{D}^{\,T} & m_{RR}
 ^{2}+m_{B}^{2}
 \end{bmatrix}.
\label{realsnu2}
\end{equation} 
where 
$m_{LL}^{2} = m_{\tilde{l}_{L}}^{2} + \dfrac{v_{u}^{2}}{2}|Y_{\nu}|^{2} + 
\dfrac{m_{Z}^{2}}{2}\cos2\beta$, $ m_{RL}^{2} = -\mu^{*}\dfrac{v_{d}}
{\sqrt{2}}Y_{\nu} + \dfrac{v_{u}}{\sqrt{2}}T_{\nu}$ and $ m_{RR}^{2} = M_{N}
^{2} + m_{N}^{2} + \dfrac{v_{u}^{2}}{2}|Y_{\nu}|^{2}$  are the 3$\times$3 
mass matrices in the flavour basis, $m_{D}\,=\,Y_{\nu}\dfrac{v_{u}}
{\sqrt{2}}$ is the Dirac mass matrix for neutrinos and $m_{\tilde{l}_{L}}$ 
is the soft mass term for the left-chiral slepton doublet $\tilde{L}$ 
(suppressing generation indices). $T_{\nu}$ will be parametrised 
as $T_{\nu}\,=\,A_{\nu}\,Y_{\nu}$, $A_{\nu}$ being a mass-scale related to 
the scale of SUSY-breaking.

Each of the $6\times6$ blocks of the mass matrix in 
equation~\eqref{realsnu2} can be diagonalized by an unitary transformation,
which mixes the right-chiral and left-chiral sneutrinos (separately 
in both CP-even and CP-odd sectors). The $3\times3$ mixing matrix between 
left-chiral and right-chiral sneutrinos is given by (when the off-diagonal
elements of each of the two blocks are much smaller than corresponding 
diagonal elements),
\begin{equation}
\Theta_{\tilde{\nu}}\,\approx\,2\,\left[m_{LL}^{2}-(m_{RR}^{2}\mp m_{B}^{2})\right]^{-1}\,\left[(m_{RL}^{2}\pm m_{D}M_{N})\right]^{T},
\label{eqn:rotang}
\end{equation} 
which is a multiplication of two matrices where the upper (lower) sign 
corresponds to mixing in the CP-even (CP-odd) sector. 

Since there is no mixing between the CP-Even and CP-odd sectors of the 
sneutrinos, both the lightest CP-even as well as the lightest CP-odd snutrino 
will be dark matter in this scenario\footnote{Since the only possible decay of the CP-odd lightest sneutrino 
is into the CP-even one and the corresponding amplitude is suppressed by the
squares of the neutrino Yukawa coupling, the lifetime of this CP-odd lightest 
sneutrino will always be larger than the age of the Universe. We have ensured 
this lifetime to be typically of order $\simeq\,10^{27}$ sec with proper 
choice of parameters.}. We are considering both (CP-even as well as CP-odd) 
lightest  sneutrinos to be dominantly right-handed ({\it{i.e}} small 
left-right mixing angle $(\Theta_{\snu})_{1j}$ with j=1,2,3) with masses
\begin{equation}
M_{\snu^{1}_{DM}(\snu^{2}_{DM})}\simeq\,\sqrt{m_{RR}^{2}\mp\,m_{B}^{2}}\,=\sqrt{\,M_{N}^{2} + m_{N}^{2} + \frac{v_{u}^{2}}{2}|Y_{\nu}|^{2}\mp\,m^{2}_{B}},
\label{eqn:LSPmass} 
\end{equation}  
where the upper(lower) sign is for the  CP-even(odd) dark matter $\snu^{1}_{DM}(\snu^{2}_{DM})$.

The right-handed sneutrinos($\snu_{R}$) being SM gauge singlets, their only 
interactions are governed by the mixing (denoted by $\Theta_{\snu}$) with 
the left-handed sneutrinos $\snu_{L}$ . Pertaining to such small
interaction strength, LSP $\snu_{R}$s never reaches thermal equillibrium and 
are produced from the decay of heavier superparticles (left-handed sleptons 
$\tilde{l}_{L}$, left-handed sneutrinos $\tilde{\nu}_{L}$, neutralinos $
\chi^{0}_{i}$ and charginos $\chi^{\pm}_{i}$).
Such production mechanism of non-thermal DM particles is called 
$\textit{freeze-in}$~\cite{Feng:2003xh, Feng:2004we,Hall:2009bx, 
Visinelli:2015eka, Waldstein:2017wps, Visinelli:2017qga,
Bernal:2017kxu, DEramo:2017ecx, Becker:2018rve, Biswas:2018ybc, Borah:2018gjk,
Biswas:2018aib}. 

The yield of a DM particle produced via  $\textit{freeze-in}$ is 
given by~\cite{Hall:2009bx},
\begin{equation}
Y_{\snu^{i}_{DM}}=\frac{n_{\snu^{i}_{DM}}}{s}\,\simeq\,\frac{45}{1.66\times4\pi^{4}}\,\frac{M_{Pl}}{g^{s}_{*}\sqrt{g^{\rho}_{*}}}\,\underset{all\,A}{\sum}\,\frac{g_{A}\,\Gamma^{i}_{A}}{m^{2}_{A}}
\int^{x=\infty}_{x=0}\,K_{1}(x)x^{3}dx\,,
\label{eqn:yield}
\end{equation}
where 
$n_{\snu^{i}_{DM}}$ and $s$ being the number density of the DM candidate and entropy density of 
the universe at the time of $\textit{freeze-in}$ respectively. Furthermore, 
$m_{A}$ and $g_{A}$ are the mass and degrees of freedom of the decaying 
superparticle A, whereas  $\Gamma^{i}_{A}$ is it's decay width to the $\snu^{i}_{DM}
$. All the relevant decay amplitudes and decay widths are given in 
Appendices~\ref{sec:Appendix1} and \ref{sec:Appendix2}. The ratio $\dfrac{m_{A}}{T}$  is 
define as $x$ and $g^{s,\rho}_{*}$ are the number of 
degrees of freedom at the time of $\textit{freeze-in}$ i.e. $T\,\simeq
\,m_{A}$ for entropy $s$ and energy density $\rho$ respectively.

The relic density via $\textit{freeze-in}$ is given by\cite{Gondolo:1997km},
\begin{equation}
\Omega_{FI}\,h^{2}=2.755\times10^{8}\,\underset{i}{\sum} M_{\snu^{i}_{DM}}\,Y_{\snu^{i}_{DM}},
\label{eqn:FIrelic}
\end{equation}
where  $M_{\snu^{i}_{DM}}$ is the mass of sneutrino dark matter and the 
index $i$ takes care of both the CP-even and CP-odd sneutrino mass 
eigenstates. Using the values of $M_{Pl}\,=\,10^{19}\,$GeV, $g^{s}_{*}\,
\simeq\,g^{\rho}_{*}\,\simeq\,100$~\cite{Kolb:1990vq} in 
equation~\ref{eqn:yield}, the $\textit{freeze-in}$ relic density of 
$\snu_{DM}$, denoted by $\Omega_{FI}$ in equation~\eqref{eqn:FIrelic} can be 
estimated as,
\begin{eqnarray}
  \Omega_{FI}\,h^{2} & \simeq & 1.92\times10^{23}\,\underset{i}{\sum}M_{\tilde{\nu}^{i}_{DM}}\,\frac{\Gamma_{tot}^i}{m_{susy}^{2}}
\label{eqn:FIrelicestimate}
\end{eqnarray}
where $\Gamma_{tot}^i\,=\,\underset{all\,A}{\sum}\,g_{A}\,
\Gamma^{i}_{A}\int^{x=\infty}_{x=0}\,K_{1}(x)x^{3}dx\,$ with  $g_{A}$ being the 
appropriate degree of freedom for each of the decaying 
particles. $M_{\snu^{i}_{DM}}$ is the mass of $\snu^{i}_{DM}$ 
and $m_{susy}$ is the generic value of the SUSY-breaking mass parameters in 
the electroweak sector of the MSSM part of the spectrum.

Over and above $\textit{freeze-in}$, the post $\textit{freeze-out}$ decay
of next-to-lightest superparticle(NLSP) also gives rise to some amount of DM 
relic, given by~\cite{Feng:2003xh, Feng:2004we}
\begin{equation}
\Omega_{FO}\,h^{2}\,=\,\left(\underset{i}{\sum}BR\,({\rm NLSP}\rightarrow \snu^{i}_{DM}+{\rm SM})\frac{M_{\snu^{i}_{DM}}}{M_{\rm NLSP}}\right)\,\Omega_{NLSP}\,h^{2},
\label{eqn:FOrelic}
\end{equation}
In our analysis we have considered the possibilities of dominantly 
right-handed stau ($\tilde{\tau}_{R}$) and bino or higgsino dominated 
neutralino($\chi^{0}_{1}$) being the NLSP.
In any case the total relic density is determined by,
\begin{equation}
\Omega_{\snu_{DM}}\,h^{2}\,=\,\Omega_{FI}\,h^{2}+\Omega_{FO}\,h^{2}.
\end{equation}
While carrying out our analysis we have ensured that our chosen parameter
region do satify the following constraints.
\begin{itemize}
\item PLANCK collaboration sets an upper limit on the DM relic 
density~\cite{Ade:2015xua},
\begin{equation}
\Omega_{CDM}h^{2}\,= 0.1199\,\pm\,0.0027.
\end{equation}
which we have respected through out our analysis in determining the allowed 
parameter space of the model.  
\item The usual constraints on neutrino masses and mixing angles, 
mostly derived from oscillation data, have been satisfied. In addition,
since neutrino Yukawa couplings($Y_{\nu}$) and neutrino Majorana 
masses($M_{N}$) play a major role in determining the left-right mixing 
angles ($\Theta_{\snu}$), all the associated bounds 
(BBN constraints on heavy neutrino decays, X-ray and Gamma-ray observations, 
DW production, Lyman-$\alpha$ forest) on the light sterile neutrinos have 
been taken into account~\cite{Adhikari:2016bei, Chan:2016aab, Hofmann:2016urz, Perez:2016tcq,Essig:2013goa,Dodelson:1993je,Baur:2017stq, Yeche:2017upn, Schneider:2016uqi}. 
\item The only possible decays of $\textit{frozen-out}$ NLSP into 
$\snu_{DM}$ being suppressed by left-right mixing $\Theta_{\snu\,1j}$,
(where j=1,2,3) the NLSPs are fairly long-lived and can potentially affect 
the standard BBN predictions~\cite{Poulin:2015opa,Kawasaki:2004qu,Jedamzik:2006xz,Banerjee:2016uyt,Cyburt:2015mya,Iocco:2008va,Ishiwata:2009gs}.
We have taken note of these constraints through out our analysis by 
assuming $\tau_{NLSP} \leq 100$ sec. It is difficult to accommodate a
bino dominated neutralino as NLSP in such scenarios since the late  
decay of such a neutralino gives rise to an unacceptably large sneutrino 
relic density. One's best bet thus is either a higgsino dominated neutralino 
or a dominantly right-handed stau to be the NLSP. In the range
$M_{NLSP}\,\simeq\,300-900\,$GeV such NLSP can only give rise to $1\%$ of
the total relic density (for $M_{\snu_{DM}}\,\simeq\,100\,$GeV) and hence 
one can safely ignore their contribution and concentrate in the 
{\it{freeze-in}} contribution solely.
\item The mass of lightest higgs emerging from the spectrum calculation using $\mathtt{SPheno}$~\cite{Porod:2011nf}
is kept in the range $123\,\text{GeV}\,<\,m_{h^{0}}\,<\,128\,\text{GeV}$~\cite{Aaboud:2018wps, Aad:2015zhl}. 
The stop mass parameters ($m_{\tilde{t}_{1,2}}$) and trilinear 
($A_{t}$) couplings have been fixed at such values as to ensure the 
observed value of $m_{h^{0}}$(see table~\ref{tab:variables}), which does 
not affect our arguments presented in this paper. The mass window have been 
kept a little lenient, in order to account for various theoretical 
uncertainties. Moreover, the consistency of lightest higgs couplings
with LHC data~\cite{Khachatryan:2014jba,Aad:2015gba} have been ensured with the help of $\mathtt{LILITH}$~\cite{Bernon:2015hsa} and $\mathtt{HiggsBounds}$~\cite{Bechtle:2013wla}.
\item The PMNS-driven contributions to FCNC processes like $\mu\,\rightarrow
\,e\,\gamma$,$\,\tau\,\rightarrow\,\mu\,\gamma$ etc. are within the current 
limits~\cite{Antusch:2014woa}.
\end{itemize}

\section{Strategy of our analysis}\label{sec:strategy}

\begin{table}[t]
\begin{center}
\begin{tabular}{|c|c|c|}
\hline
Parameter  & Ranges($m_{susy}\simeq\,1\,$TeV) & Ranges($m_{susy}\simeq\,100\,$TeV) \\\hline\hline
$\{\theta_{12}\,,\theta_{23}\,,\theta_{13}\,\}$& $\{33.62^\circ,47.2^\circ,8.54^\circ\}$ & $\{33.62^\circ,47.2^\circ,8.54^\circ\}$      \\
$\{\delta_{CP}\,,\alpha_{1}\,,\alpha_{2}\}$  & $\{\frac{3\pi}{2},\frac{\pi}{4},\frac{\pi}{4}\}$ & $\{\frac{3\pi}{2},\frac{\pi}{4},\frac{\pi}{4}\}$\\
$(M_{N})_{11}$  & 10 keV & 10 keV \\
$M^{H}_{N}$     & 10 keV - 50 keV, 500 MeV - $\mathcal{O}$(10 GeV) & $\mathcal{O}$(10 GeV)-200 GeV  \\
$(m_{N})_{11}$ & 20 GeV - 400 GeV & 20 GeV - 200 GeV \\
$(m_{B})_{11}$  & 10 GeV  & 10 GeV  \\
$\mu$           & 500 GeV - 1.5 TeV & 500 GeV - 1.5 TeV \\
$\tan\beta$     & 10.5  & 3.5 \\
$A_{\nu}$       & 20 GeV - 400 GeV & 20 GeV - 400 GeV\\
$(m_{N})^{H}$  & 1 TeV & 1 TeV \\
$(m_{B})^{H}$   & 175 GeV & 175 GeV \\\hline\hline
$m_{\tilde{t}_{1}}$ &  1225 GeV &  500 TeV \\
$m_{\tilde{t}_{2}}$ & 1725 GeV &  502.4 TeV \\
$A_{t}$             & 3890 GeV &  2089 TeV \\
$m_{h^{0}}$         & 125.62 GeV  &  124.77 GeV \\
$m_{A^{0}}$         & 2 TeV &  3640 TeV   \\
$m_{\tilde{q}_{L,R}}$ & 3 TeV & 600 TeV \\
$m_{\tilde{g}}$       & 3 TeV & 600 TeV \\
\hline
\end{tabular}
\caption{Input parameters and their ranges scanned in our analysis are 
tabulated here. The neutrino oscillation parameters, $(m_{N})^{H}$, 
$(m_{B})^{H}$, $(M_{N})_{11}$ and $(m_{B})_{11}$ have been kept fixed 
as they have very little or no effect on the analysis carried out. The 
Majorana masses $(M_{N})^{H}$ and other supersymmetric parameters like 
$(m_{N})_{11}$, $\mu$, $\tan\beta$ and $A_{\nu}$
have been varied such that one can satisfy the necesssary constraints 
on the $\textit{freeze-in}$ relic density 
(see equations.~\eqref{eqn:FIrelicbound} and ~\eqref{eqn:highmsusyFIrelicbound} respecively).}
\label{tab:variables}
\end{center}
\end{table} 

We summarize next the methodology adopted in obtaining the {\it new results}
contained in this paper.
The sneutrino relic density via $\textit{freeze-in}$ depends on the 
decay width of heavier superparticles into it. Hence the corresponding decay 
widths and the left-right mixing angles $\Theta_{\snu\,1j}$ (with j=1,2,3)
are highly constrained by the PLANCK limit of relic abundance~\footnote{Henceforth we will denote all the mixing angles of $\snu$ LSP
collectively as $\Theta_{\snu\,1j}$ without mentioning the values j can
acquire since our arguments apply to all possible j values. When we refer to 
any particular mixing angle the specific value of j will be mentioned.}. 
For a dominantly right-handed sneutrino dark 
matter, all the mixing angles $\Theta_{\snu\,1j}$ depend on $m_{susy}\approx 
m_{\tilde{l}_{L}}$. The bino ($m_{\tilde{B}}$) and wino 
($m_{\tilde{W}^{3}}$) mass parameters are also in the
same value for simplicity, and henceforth $m_{\tilde{l}_{L}}\,,m_{\tilde{B}}
$ and $m_{\tilde{W}^{3}}$ will collectively be denoted as $m_{susy}$.

We have considered two representative cases, with $m_{susy}\,\simeq\,1\,$TeV 
 (which is accessible at the LHC) and $m_{susy}\,\simeq\,\mathcal{O}(100)\,
$TeV (inaccessible at the colliders). For any $m_{susy}$, equation~\eqref{eqn:rotang} and \eqref{eqn:FIrelicestimate} allow one to translate 
the limits on  $\Theta_{\snu\,1j}$ into constraints on the parameters 
$A_{\nu}$, $\mu$, $\tan\beta$, $Y_{\nu}$ and, last but not the least, on 
$m_{RR}$ and $m_{B}$ (see section~\ref{sec:details1} for details).

In order to study the 
dependence of relic density on these parameters, we have varied them in 
several ranges quoted in table~\ref{tab:variables}. 
The best fit values for the neutrino oscillation parameters 
have been taken from~\cite{Esteban:2016qun}. The Majorana phases ($\alpha_{1}
\,,\alpha_{2}$) have  been set to $\frac{\pi}{4}$ since variation in them 
does not change the order of magnitude of $Y_{\nu}$, which is all important for 
the scenario under consideration. $(M_{N})_{11}$, $(m_{N})_{11}$
($(m_{B})_{11}$) are respectively the Majorana mass, $\Delta\,L\,=0(2)$ 
SUSY-breaking masses for the LSP sneutrino, while $M_{N}^{H}$, $(m_{N})^{H}$
($(m_{B})^{H}$) are the corresponding parameters for the remaining two 
heavier right-handed sneutrinos. One should note that in both of the 
aforementioned cases of $m_{susy}$, we have used 
$(m_{B})_{11}\,<<\,(m_{N})_{11}$ in order to ensure 
$\tau_{\snu^{2}_{DM}}\,\geq\,10^{27}\,$sec. For 
$m_{susy}\,\simeq\,1\,$TeV, the region $50\,{\rm keV}\,<M^{H}_{N}\,<500\,
{\rm MeV}$ has been left out, since such values cause the heavy neutrinos 
to decay after 100 sec.~\cite{Shakya:2015xnx}, thereby disturbing standard 
BBN predictions. Overall, while our analysis is insensitive to the
choices of $(m_{N})^{H}$ and $(m_{B})^{H}$, the benchmark values (shown in
table~\ref{tab:variables}) are chosen to ensure that the heavier sneutrino
mass eigenstates are heavier than the NLSP and decay before BBN. The same 
argument applies to the parameters ($m_{\tilde{t}_{1,2}}\,,A_{t}\,,m_{\tilde{q}_{L,R}}\,,m_{\tilde{g}}\,,m_{A^{0}}$) 
related to strongly interacting sector and higgs sector of the model. 
All these parameters have been kept fixed through out the analysis at the 
values quoted in table~\ref{tab:variables} although our conclusion is not 
specific to such values.

{\bf Higgsino freeze-in :} The situation is different when the higgsino 
component in $\chi_i^0$ ($i$=1,..4) decays into $\snu^{i}_{DM}$ while the 
former is in the equilibrium with the thermal bath. The decay amplitude  in 
this case depends solely on $Y_\nu$. If the right sneutrino LSP is of the 
same family as that of the lightest active neutrino, then the higgsino decay 
width to the LSP sneutrino is solely determined by the lightest neutrino 
Yukawa couplings $(Y_\nu)_{1i}, (i=1,2,3)$, assuming 
normal hierarchy(NH) of neutrino masses~\footnote{For inverted hierarchy 
(IH), the conclusions are not very different if the sneutrino associated 
with the lightest active neutrino is the DM candidate. For the quasi-
degenerate scenario, the lightest neutrino mass is bound to be $\simeq\,
10^{-2}\,\text{eV}$. In order to prevent higgsino decays from overproducing 
sneutrino dark matter via Yukawa interaction, one has to lower the Majorana 
mass corresponding to the lightest neutrino eigenstate. One thus finds 
oneself pushed down to eV-scale value for this Majorana mass. Such a 
situation has already been mentioned in~\cite{Gopalakrishna:2006kr}; 
however, recent limits from BBN and recombination
~\cite{Cyburt:2015mya, Ade:2015xua} strongly restricts this scenario.}. 
We have found that for $(Y_{\nu})_{1i}\,\simeq\,10^{-12}$ the higgsino
decay saturates the sneutrino relic abundance. Since the lightest active 
neutrino mass is a free parameter in the hierarchical scenario, it can be 
tuned to control the contribution of any higgsino-dominated neutralino to 
relic abundance. Using this freedom if we reduce these Yukawa couplings to 
 $\mathcal{O}(10^{-14})$ or less, the higgsino contribution is even lesser.
In order to look for the effect of other parameters on the relic density 
we have been a little indulgent in our choice of $(Y_{\nu})_{1i}$.
Throughout our analysis, these Yukawa couplings  corresponding to the 
flavour eigenstates dominating the lightest mass eigenstate
is thus set to $\,\mathcal{O}(10^{-13})$ thus yielding the 
lightest neutrino mass $m_1 \simeq 10^{-8}$ eV for a 10 keV Majorana mass. 
The {\it{freeze-in}} of Higgsinos in this case contributes about 
$1\%$ of the total relic of $\snu_{DM}$.
This choice considerably simplifies our analysis. The more general mixing 
scenario multiplies the number of free parameters but does not affect the 
constraint on the Majorana mass of the lightest neutrino. 
The Yukawa couplings for other neutrinos are dictated by the Majorana 
mass(es) and neutrino oscillation data. 
These couplings can be considerably large compared to the pure Dirac 
case. One can use the Casas-Ibarra parametrization~\cite{Casas:2001sr}\footnote{Since for our present analysis only the 
order of magnitude of the Dirac Yukawa matrix elements matters, the Casas-
Ibarra parametrization helps us fix the Yukawa matrix conveniently which 
satisfies neutrino oscillation data by construction, without any random 
sampling.} to write,
\begin{equation}
Y_{\nu} = \frac{-i \sqrt{2}}{v_{u}}\,\sqrt{M_{N}}\,R\,\sqrt{m_{\nu}^{diag}}\,U_{PMNS}^{\dagger}.
\label{eqn:Yukawa}
\end{equation}
where, $R$ is a complex orthogonal matrix, $m^{diag}_{\nu}$ is the 
diagonalized light-neutrino mass matrix (whose elements are determined by 
$\Delta\,m^{2}_{\odot}$ and $\Delta\,m^{2}_{\oplus}$) and $U_{PMNS}$ is the 
diagonalizing matrix. Since one light neutrino is  effectively massless, we 
have parametrized $R$ as~\cite{Abada:2006ea}
\begin{center}
R =
$\begin{bmatrix}
1 &&    0         &&    0 \\
0 &&  \cos\,\omega && \sin\,\omega \\
0 && -\sin\,\omega && \cos\,\omega
\end{bmatrix}$,
\end{center}
where a non-vanishing $R_{11}$ with appropriate $m_{\nu}^{diag}$ reflects 
the postulated hierarchy. We use $\omega\,=\,i$.

\section{Results}\label{sec:details1}
As has been mentioned in section~\ref{sec:strategy}, very high $\Delta\,L=2$
neutrino masses lead to Yukawa couplings that are inadmissible from 
$\textit{freeze-in}$ constraints. We consider two representative values of
$m_{susy}$, namely 1 TeV and 100 TeV. Both of these ranges are consistent 
with $\textit{freeze-in}$ constraints. 

\subsection{Results for $m_{susy}\,\simeq\,1\,$TeV}
\label{sec:lowsusy}

In this scenario with all MSSM superparticles in the TeV scale, equation~\eqref{eqn:FIrelicestimate} implies~\cite{Ade:2015xua},
  
\begin{eqnarray}
\underset{i}{\sum}\,M_{\snu^{i}_{DM}}\Gamma_{\snu^{i}_{DM}}\,\lesssim\,5.2\times 10^{-19}\,{\rm GeV^{2}}.
\label{eqn:FIrelicbound} 
\end{eqnarray}

The sfermions give the dominant contribution to the decay width 
compared to the contributions from bino and winos since 
it is enhanced by a phase-space factor of 
$\dfrac{m^{2}_{susy}}{m^{2}_{EW}}$ ({\it{e.g}} compare ~\eqref{eqn:decays1}-\eqref{eqn:decays2}). 
Thus for DM masses around 100 GeV, equation~\eqref{eqn:FIrelicbound}
leads to $(\Theta_{\snu})_{1j}\,\simeq\,10^{-12}$ in order to satisfy 
correct {\it{freeze-in}} relic density. 
Following equation~\eqref{eqn:rotang} (after neutrino mixing 
pattern is taken into account) one finds that,
\begin{equation}
\left[Y_{\nu}\left((A_{\nu}-\mu \cot\beta)\mathbb{I}\,\mp M_{N}\right)\right]_{ij}\,\simeq\,10^{-8}\,{\rm GeV}\,\,{\rm for\,\,i,j=1,2,3}, 
\label{eqn:thetanumcons}
\end{equation}
where $\mathbb{I}$ is $3\times3$ identity matrix.

In the absence of $\Delta\,L\,=\,2$ masses ($M_{N}$), 
equation~\eqref{eqn:thetanumcons} is easily satisfied. In this 
case neutrino oscillation data fixes $Y_{\nu}\,\simeq\,10^{-13}$, and hence
one needs $(A_{\nu}-\mu \cot\beta)\,\leq\,10^{5}\,{\rm GeV}$, which is
consistent with $A_{\nu}$ and $\mu$ on the order of several hundreds of GeV. 
On the other hand, the presence of $M_{N}$ further constrains $Y_{\nu}
$, thus satisfying equation~\eqref{eqn:thetanumcons} for $(A_{\nu}-\mu\cot
\beta)$ within a limited range. This allowed range depends on 
$M_{\snu^{i}_{DM}}$, with larger values of $M_{\snu^{i}_{DM}}$ allowing 
smaller $(A_{\nu}-\mu\cot\beta)$ following equation~\eqref{eqn:FIrelicbound}.
In determining this allowed range we have taken into consideration 
the relic density of both the CP-even and CP-odd LSP states i.e. in our
notation $\Omega_{FI}h^{2}\,=\,\Omega_{\snu^{1}_{DM}}h^{2}+\Omega_{\snu^{2}_{DM}}h^{2}$. 
Following the equation~\eqref{eqn:thetanumcons} it is also evident
that the matrix $Y_{\nu}M_{N}$ is also tightly constrained. As already 
mentioned in section~\ref{sec:strategy}, the Yukawa couplings corresponding
to lightest neutrino ($(Y_{\nu})_{1i}$) have been chosen to be very small
and hence the corresponding Majorana mass $(M_{N})_{11}$ is not so 
tightly constrained as the other Majorana masses $(M_{N})_{H}$. We kept
$(M_{N})_{11}$ fixed during our analysis.
On the other hand, Majorana masses ($M^{H}_{N}$) are allowed only upto 
a few tens of GeV. 
In view of the BBN constraints on the lifetime of heavy neutrinos,
we have considered two possible
ranges of $M^{H}_{N}$:
\begin{enumerate}
\item In the first case we considered all the Majorana masses($(M_N)_{11}\,,M_N^H$) to be on the keV-scale and denote the Majorana mass matrix 
as $M_{N}\,\mathbb{I}$, where $M_{N}\,=(M_{N})_{11}\,=M_{N}^{H}$ . 
\item In the second case, the heavier Majorana masses 
($(M_{N})_{22,33}\,=\,M_N^H$) have been assumed to be in the range  
500 MeV - 20 GeV while the lightest $\Delta\,L\,=2$ mass $(M_{N})_{11}$
is kept fixed at 10 keV.
\end{enumerate}

\subsubsection{keV scale Majorana mass}
\label{sec:keVmajo}

\begin{figure}[t]
\begin{center}
\includegraphics[width=5.1cm,angle=270]{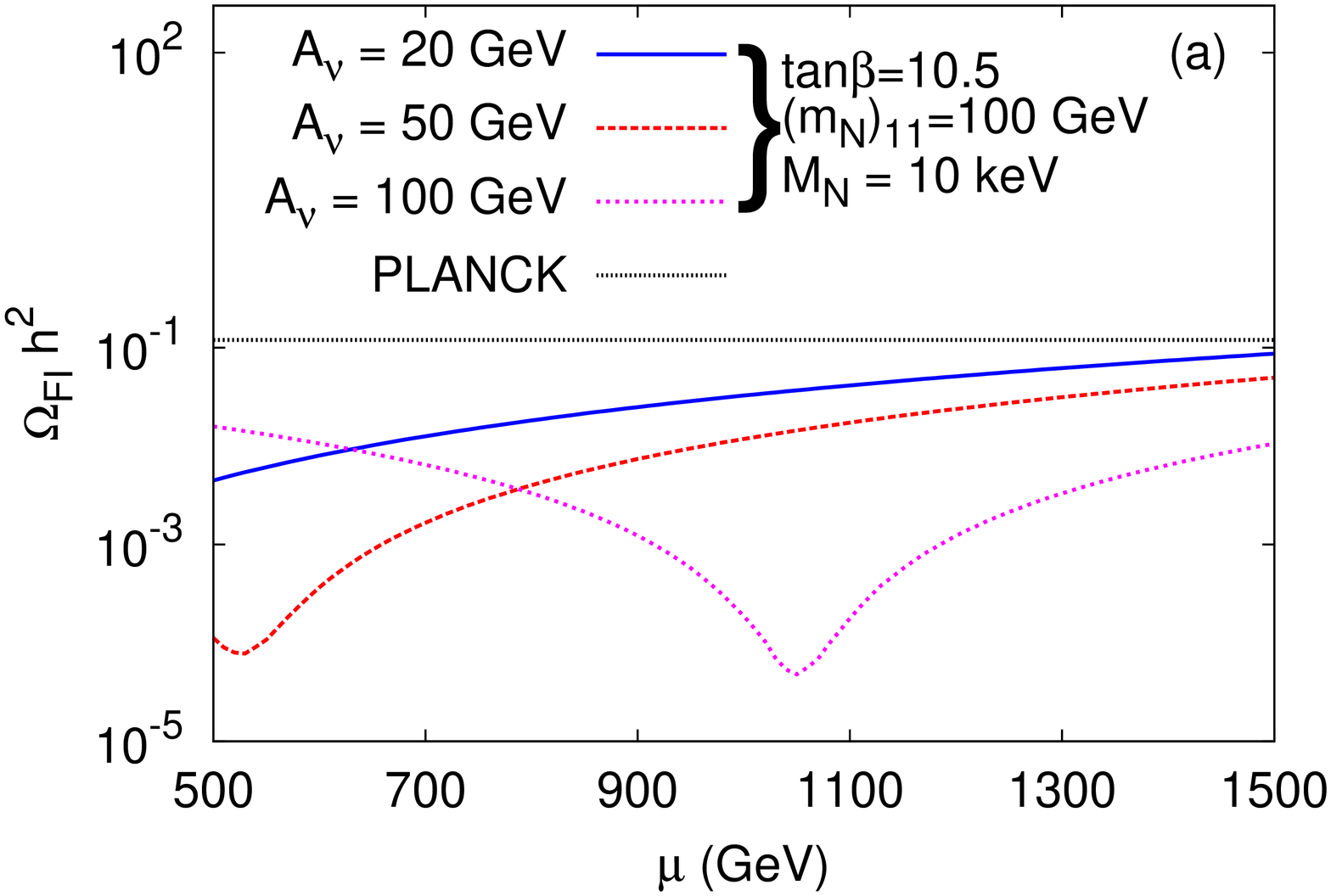}
\includegraphics[width=5.1cm,angle=270]{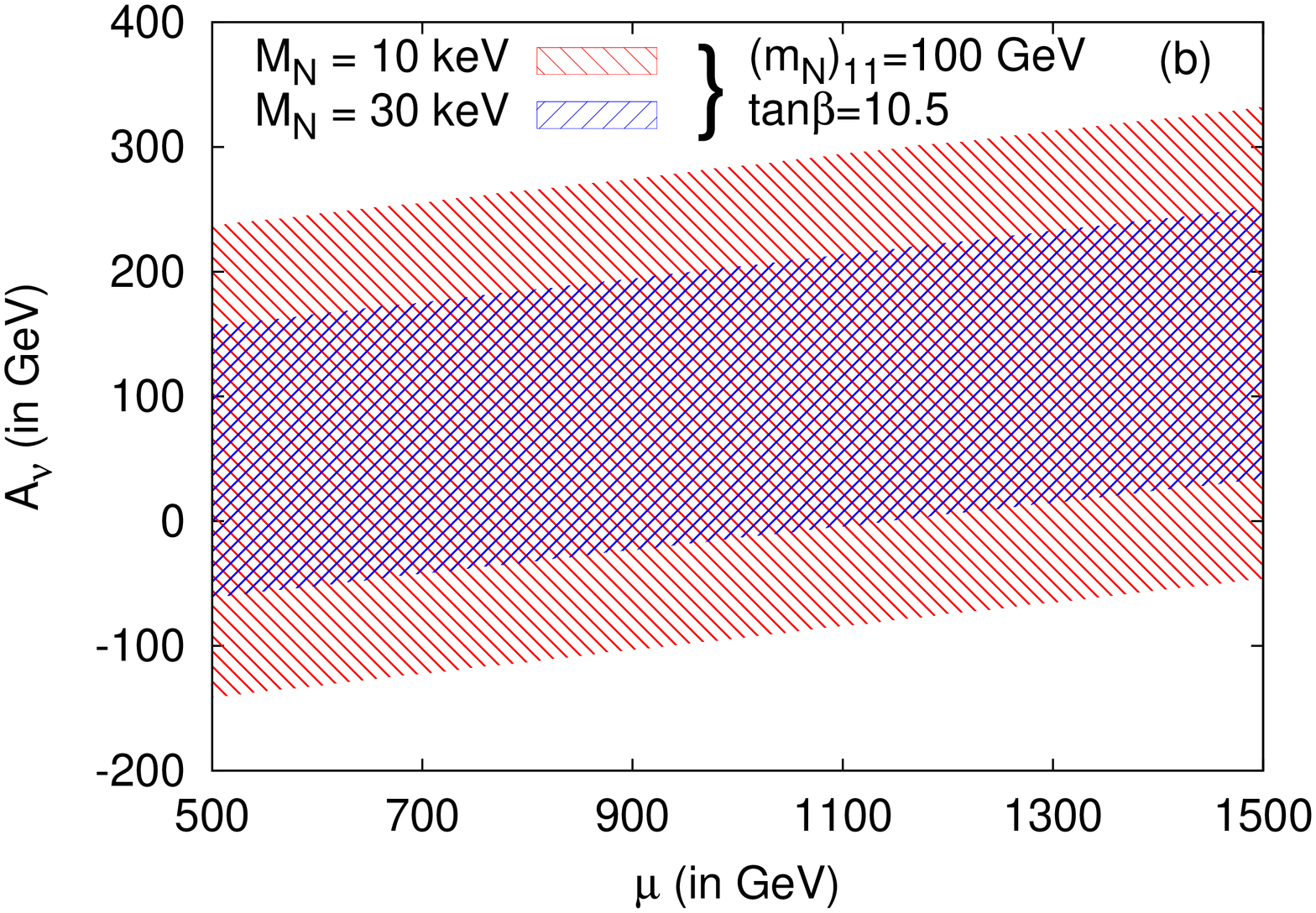}
\includegraphics[width=5.1cm,angle=270]{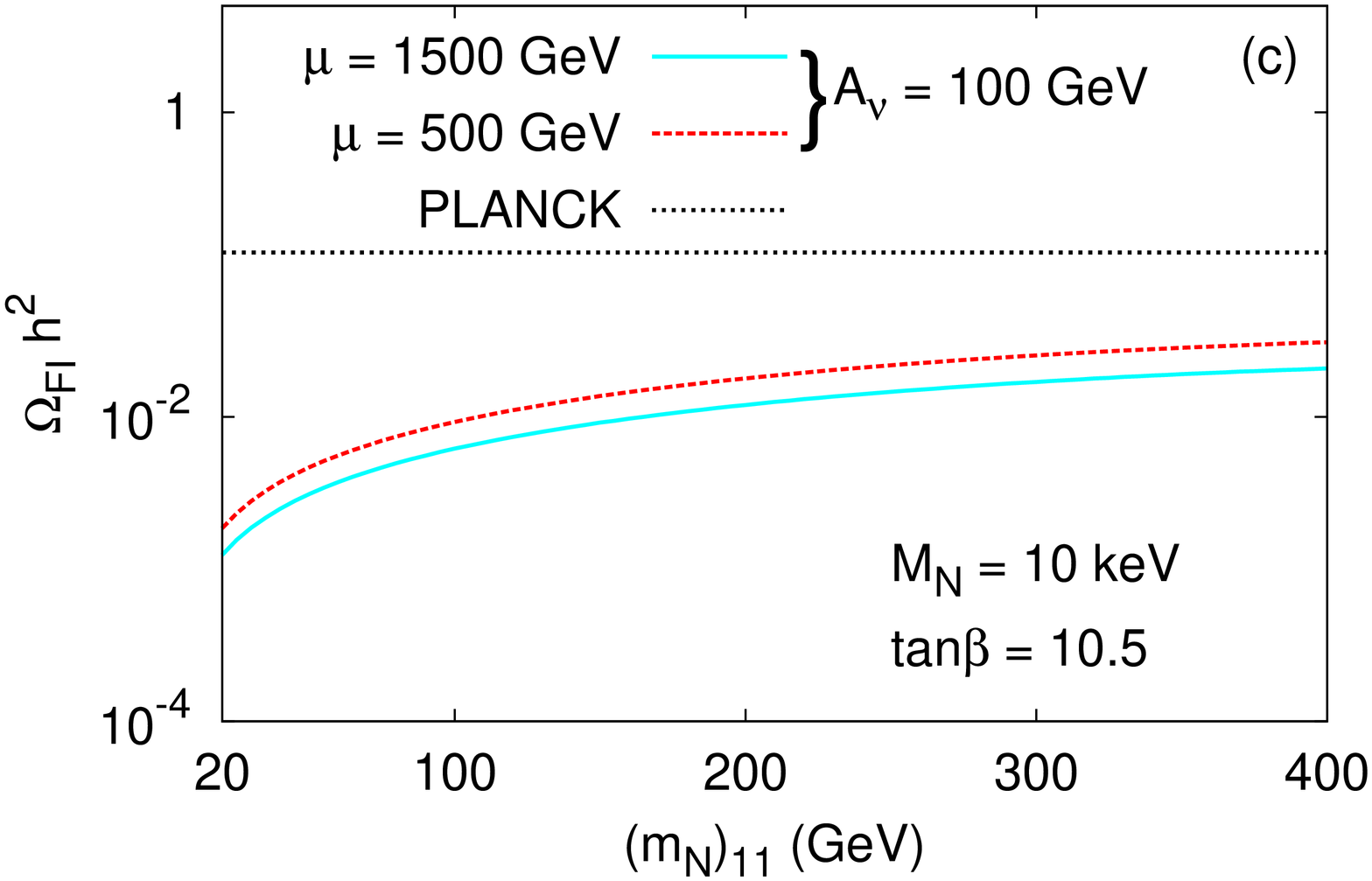}
\includegraphics[width=5.1cm,angle=270]{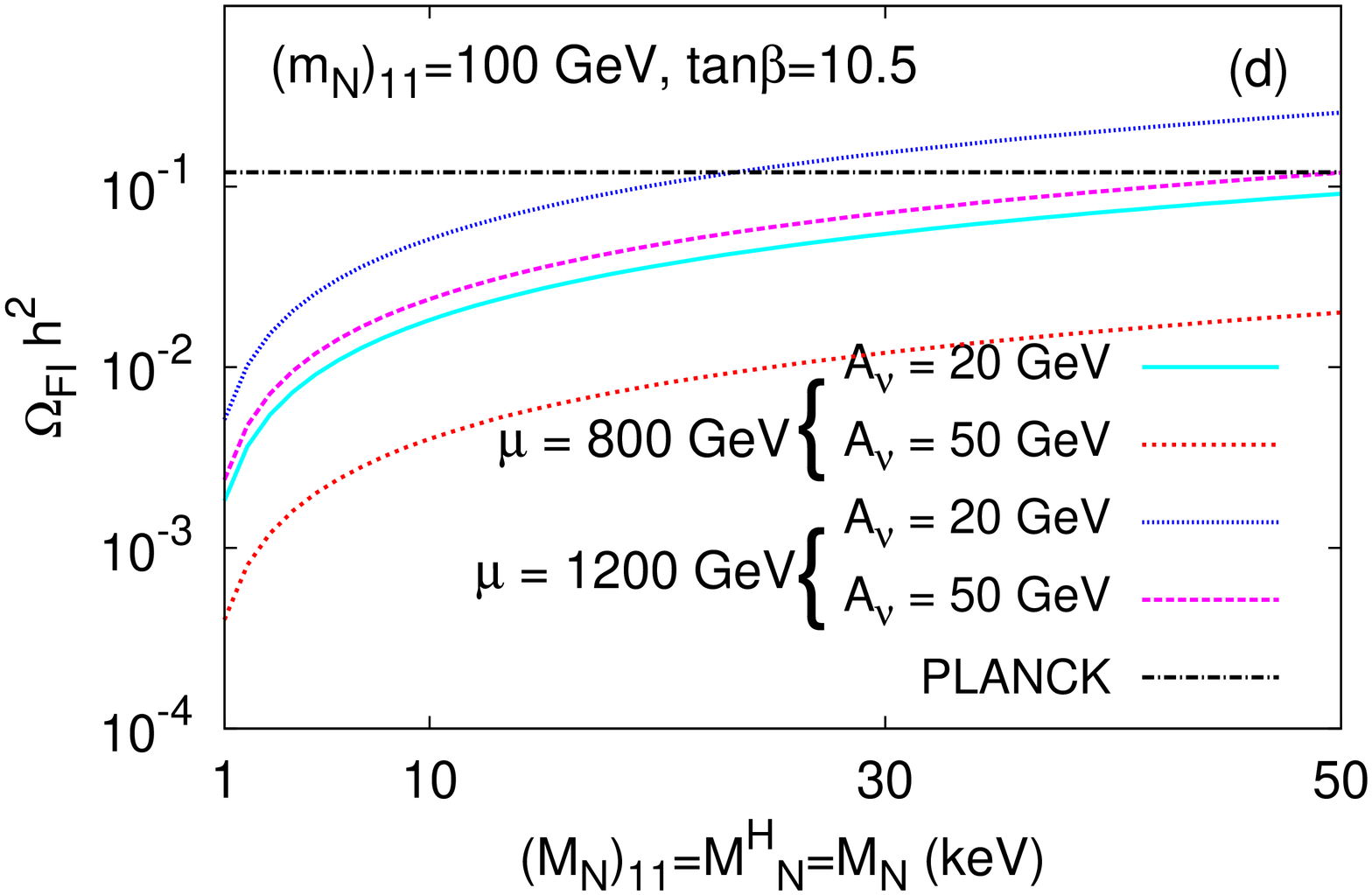}
  \caption{For $m_{susy}\,=\,1$TeV the dependence of freeze-in relic 
density with $\mu$ (subfigure~\ref{fig:paramdepkeVmajorana}(a)), 
$(m_{N})_{11}$ (subfigure~\ref{fig:paramdepkeVmajorana}(c)) and 
$M_{N}$ (subfigure~\ref{fig:paramdepkeVmajorana}(d)) have been shown. 
All the Majorana masses have been taken in the keV scale, 
$A_{\nu}$ is varied in the EW scale and $\tan\beta\,=10.5$ has been taken.
The allowed region in the $\mu-A_{\nu}$ plane for $M_{N}\,=\,10\,{\rm keV}$
is shown in the subfigure~\ref{fig:paramdepkeVmajorana}(b). 
This plot also embodies how the allowed region of $\mu-A_{\nu}$ plane 
shrinks as $M_{N}$ is increased.}
\label{fig:paramdepkeVmajorana}
\end{center}  
\end{figure}

The dependence of relic density on the supersymmetric parameters 
($A_{\nu}$, $\mu$, $(m_{N})_{11}$) and Majorana masses($M_{N}$) are shown in 
figure~\ref{fig:paramdepkeVmajorana}. As argued in section~\ref{sec:strategy}, the determining factor are the mixing angles 
$\Theta_{\snu\,1j}$ and hence the elements of the matrix $Y_{\nu}(A_{\nu}-\mu\cot\beta\,\mathbb{I}\,\mp\,M_{N}\mathbb{I})$. 
For all the elements of the matrix $M_{N}$ are in the keV range, the largest 
element of the matrix $Y_{\nu}$ is $\simeq\,10^{-10}$, leading to $M_{N}
(Y_{\nu})_{ij}\,\simeq\,10^{-14}\,{\rm GeV}$ for all i,j=1,2,3, which is
much smaller than $10^{-8}\,$GeV that is required for correct relic-density
(see equation~\eqref{eqn:thetanumcons}). Thus the correct relic density 
requires,
\begin{equation}
\left(Y_{\nu}\right)_{ij}(A_{\nu}-\mu \cot\beta)\,\simeq\,10^{-8}\,{\rm GeV}\,\,{\rm for\,\,i,j=1,2,3}, 
\label{eqn:kevthetanumcons}
\end{equation}
 which indicates  $|A_{\nu}-\mu\cot\beta|\simeq\mathcal{O}(100\,{\rm GeV})$ 
 is allowed in order to obtain correct relic density for $M_{\snu^{i}_{DM}}
\simeq\,100\,{\rm GeV}$. One may wonder how is this situation 
different from the pure Dirac neutrino masses. One should remember that 
in case of pure Dirac masses all the elements of $Y_{\nu}\,\simeq\,
\mathcal{O}(10^{-13})$, while even the tiny Majorana masses considered here 
have raised some of the elements of $Y_{\nu}$ to $\mathcal{O}(10^{-10})$ 
thereby affecting the allowed range of $|A_{\nu}-\mu\cot\beta|$.
The  figure~\ref{fig:paramdepkeVmajorana}(a) depicts how the relic
density varies with $\mu$ for given values of $A_{\nu}$, while figure~\ref{fig:paramdepkeVmajorana}(b)
shows the allowed region in $\mu-A_{\nu}$ plane for $(m_{N})_{11}\simeq\,
100\,{\rm GeV}$ and two different values of $M_{N}$. As $M_{N}$ increases 
from $10\,{\rm keV}$ to $30\,{\rm keV}$, the elements of $Y_{\nu}M_{N}$ in 
equation~\eqref{eqn:thetanumcons} increases and consequently the allowed 
region in $\mu-A_{\nu}$ plane shrinks.
 For each of these plots a minimum of relic density occurs when 
$\mu\cot\beta\,=\,A_{\nu}$. As $\mu$ departs from this point of exact 
cancellation relic density increases. In figure~\ref{fig:paramdepkeVmajorana}(c) and (d) we have showed the variation 
of relic density with $(m_{N})_{11}$ and $M_{N}$ respectively. 
With increase in $(m_{N})_{11}$, $M_{\snu^{i}_{DM}}$ increases and 
relic density being proportional to $M_{\snu^{i}_{DM}}$ (see 
equation~\eqref{eqn:FIrelicestimate}), also increases linearly with 
$(m_{N})_{11}$. Whereas, with increase in $M_{N}$, $(Y_{\nu})_{ij}$ 
increases and relic increases pertaining to increase in $\Theta_{\snu\,1j}$.  

Thus there is ample parameter space available for a right-sneutrino to be 
the LSP in presence of $M_{N}\,\simeq\,\mathcal{O}(keV)$. However, 
our choice of keV-scale Majorana masses ($M_{N}$) force all the 
sterile neutrinos to be warm DM. The Yukawa couplings($Y_{\nu}$), being 
always consistent with neutrino oscillation data, yield too large a relic
density via DW mechanism for the corresponding heavy 
neutrinos~\cite{Shakya:2015xnx} owing to their large mixing angles. 
 
\subsubsection{Majorana masses in the  MeV-GeV range}
\label{sec:MeVGeVmajo}

\begin{figure}[t]
\begin{center}
  \includegraphics[width=5.1cm,angle=270]{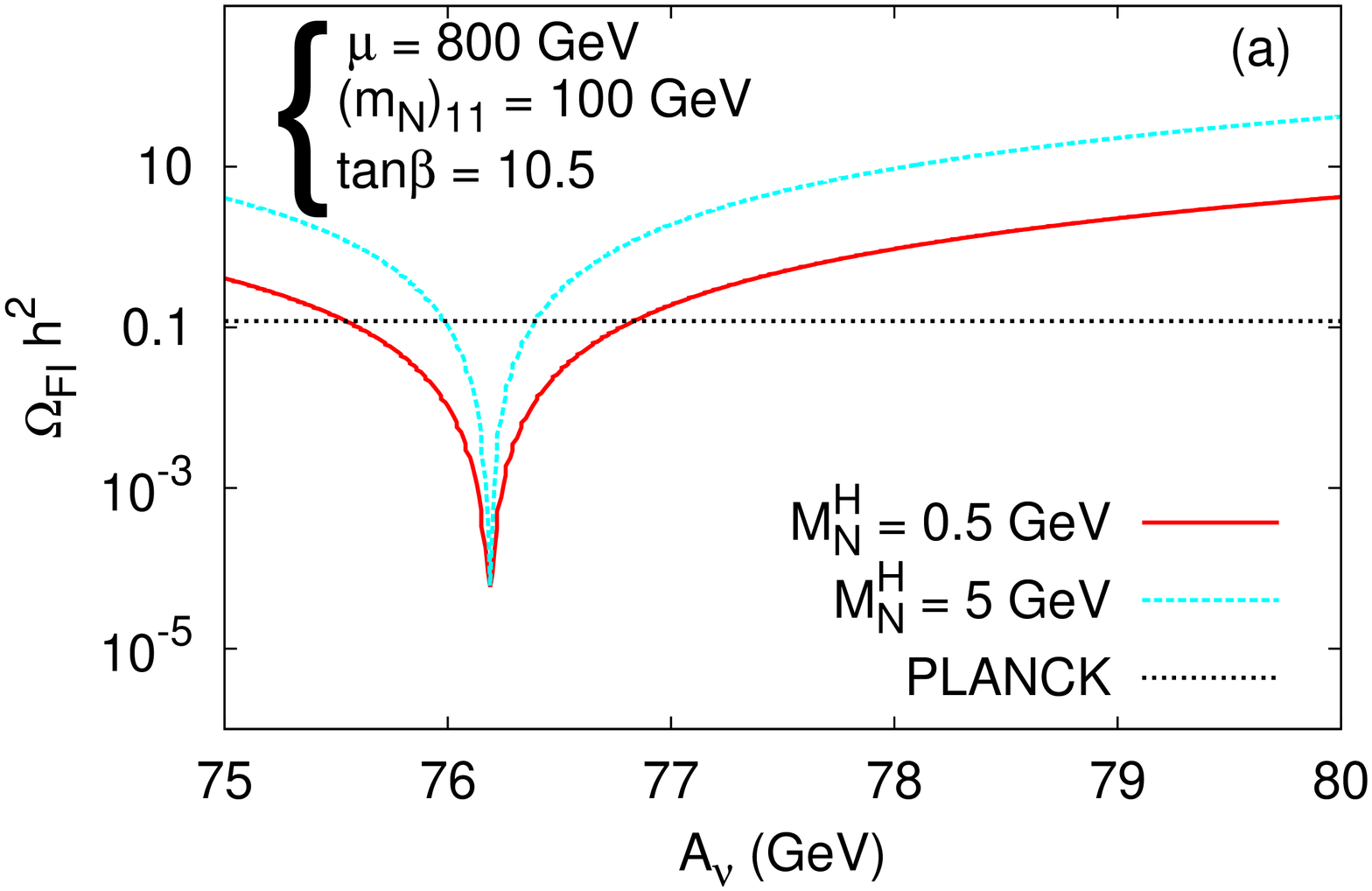}
  \includegraphics[width=4.7cm,angle=270]{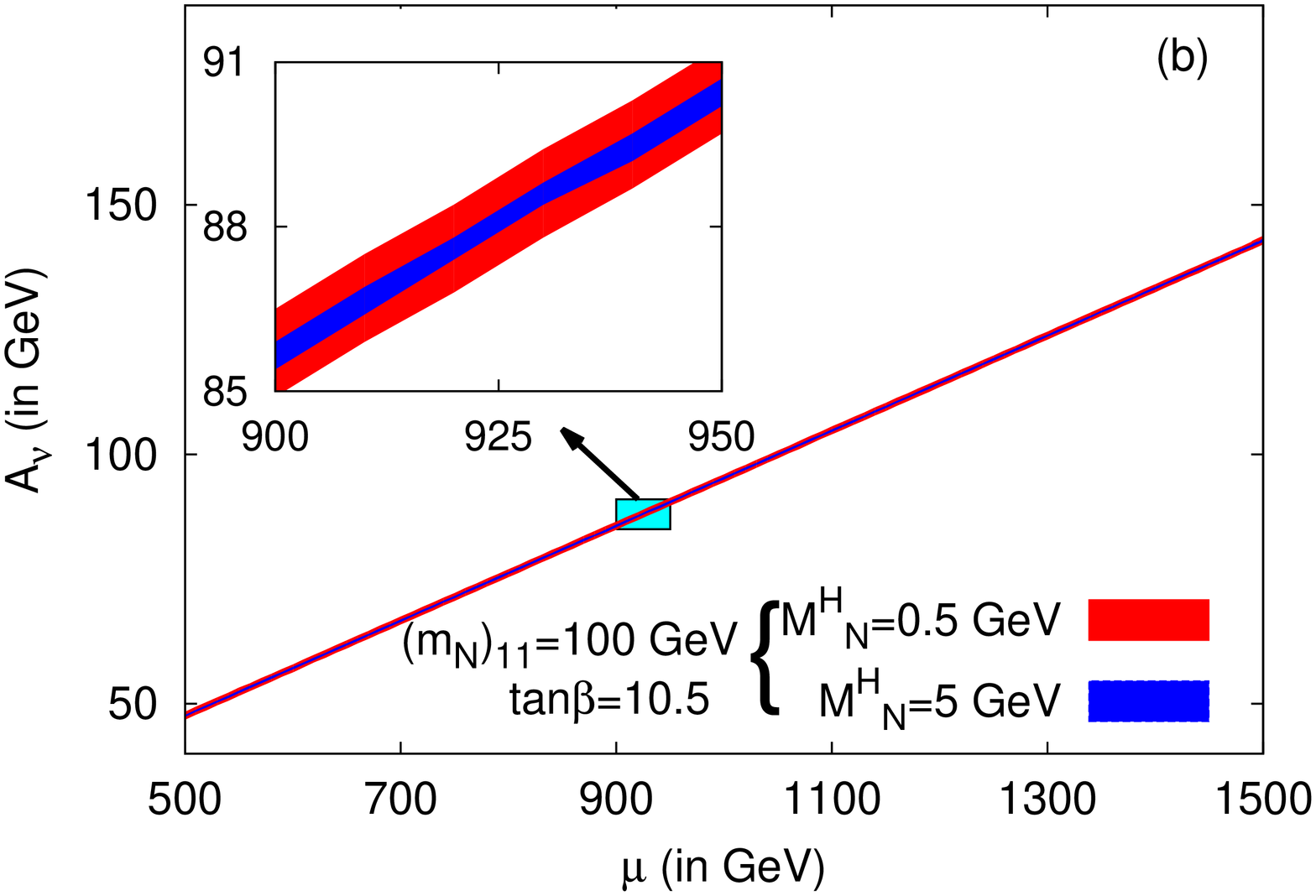}
  \includegraphics[width=7.0cm]{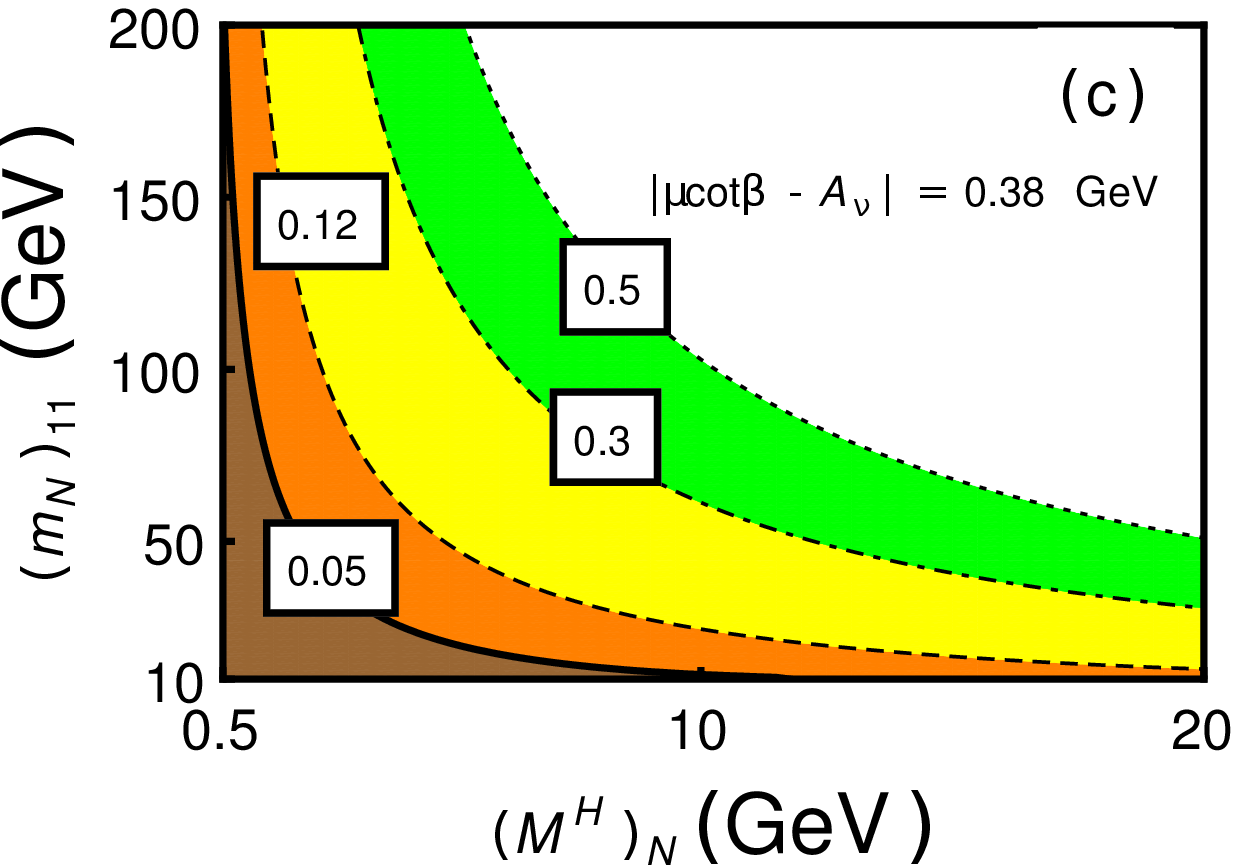}
  \includegraphics[width=7.0cm]{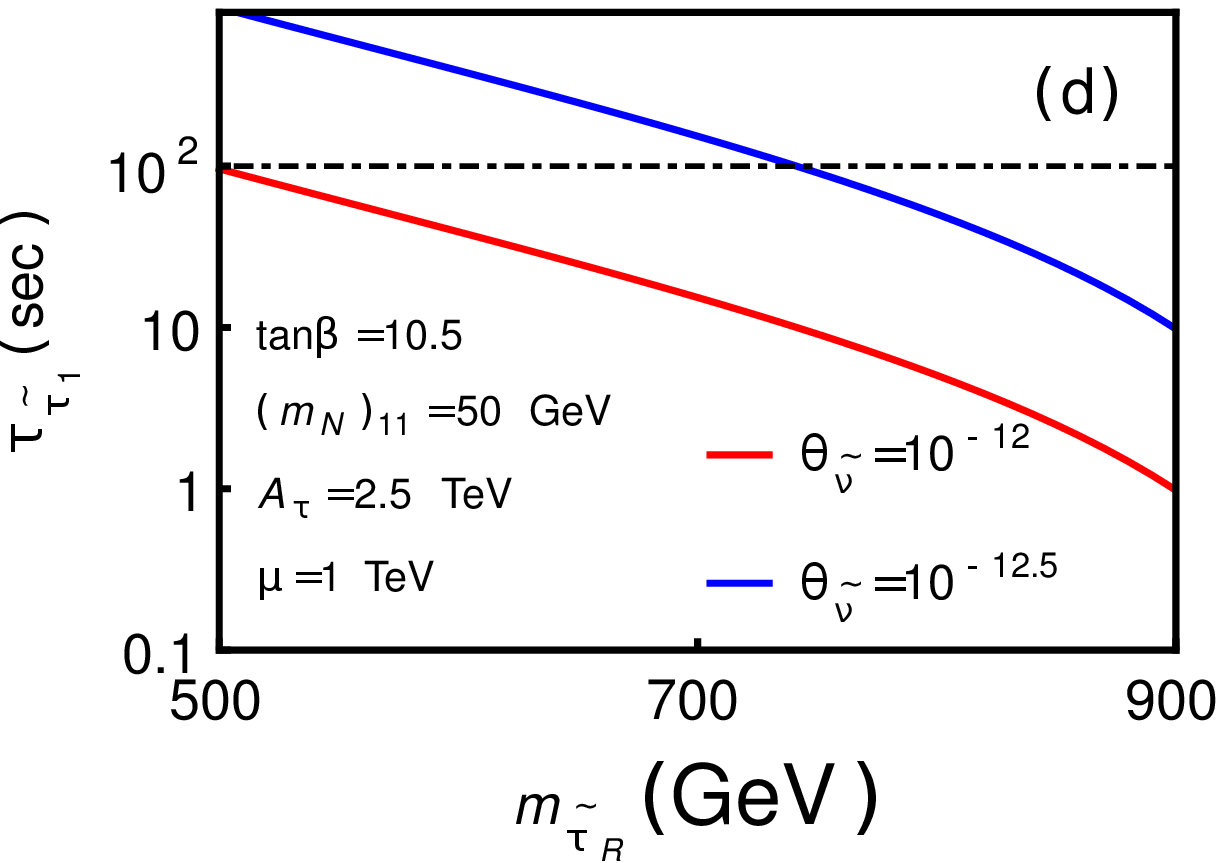}
  \caption{Subigure~\ref{fig:paramdepMeVmajorana}(a) shows how how the exact 
  cancellation between supersymmetry conserving term $\mu\,\cot\beta$ and 
  supersymmetry breaking term $A_{\nu}$ leads to correct $\textit{freeze-in}
  $ relic density for sneutrino with MeV-GeV scale Majorana masses. 
  Subfigures~\ref{fig:paramdepMeVmajorana}(b) 
  and ~\ref{fig:paramdepMeVmajorana}(c)
  show the variation of $\textit{freeze-in}$ relic density with 
  $(m_{N})_{11}$ and $M_{N}$ respectively. For each of 
  the different values of $\mu$, $A_{\nu}$ has been chosen such that 
  equation~\eqref{eqn:thetanumcons} is satisfied and one obtains the maximum 
  possible values of $(m_{N})_{11}(M_{N})$ for that choice of 
  $|A_{\nu}-\mu\cot\beta|$. If one chooses $A_{\nu}$ and $\mu\cot\beta$ to
  cancel more finely, one will get a larger value of $(m_{N})_{11}(M_{N})$,
  but the order of magnitude will not change. 
  Subfigure~\ref{fig:paramdepMeVmajorana}(d) shows the allowed 
  $\tilde{\tau}_{R}$ soft masses $m_{\tilde{\tau}_{R}}$ of
  $\tilde{\tau}_{1}$ NLSP such that it's lifetime is $\leq\,100\,$sec.
  All the parameters $(A_{\tau},\mu,\tan\beta,\theta_{\snu}\approx\,
  \Theta_{\snu\,13})$ that contributes to $\tilde{\tau}_{1}$ lifetime, are 
  quoted in the figure. We have assumed $\tan\beta\,=10.5$ in all the 
  cases.}
\label{fig:paramdepMeVmajorana}
\end{center}
\end{figure} 

Since the range of 50 keV - 500 MeV is disfavoured by BBN constraints,
we now focus on heavy neutrino Majorana masses ($M^{H}_{N}$) 
in the 500 MeV-{\it a few tens} of GeV range. 
The dependence of right sneutrino relic density on SUSY parameters 
$A_{\nu}$, $\mu$, $(m_{N})_{11}$ and $M^{H}_{N}$ are shown in 
figure~\ref{fig:paramdepMeVmajorana}. The largest value of  
$(Y_{\nu})_{ij}$ is $\simeq\,10^{-8}$ for 
$M^{H}_{N}\,\simeq\,500\,{\rm MeV}$, 
implying the largest element of $Y_{\nu}M^{H}_{N}\,\simeq\,10^{-8}\,{\rm GeV}$. 
Thus one requires $|A_{\nu}-\mu\cot\beta|\,\simeq\,1\,{\rm GeV}$  
in order to obtain the correct relic density via freeze-in. 
This cancellation is clear in Figure~\ref{fig:paramdepMeVmajorana}(a) where 
for a given $\mu$,  a very small range of $A_{\nu}$ is allowed in order to 
satisfy the relic density.  The allowed parameter space in  $\mu-A_{\nu}$ 
plane  where relic density is below the observed value, is depicted in 
figure~\ref{fig:paramdepMeVmajorana}(b) with
red and blue band for $M^{H}_{N}$ = 500 MeV and 5 GeV respectively. A 
magnified version of the region $900\,{\rm GeV}\leq\mu\leq950\,{\rm 
GeV}$ is shown in the inset of figure~\ref{fig:paramdepMeVmajorana}(b). 
Which clearly shows the decrease in the permissible values of $A_{\nu}$ for 
a given $\mu$ as one increase Majorana mass ($M^{H}_{N}$) from 500 MeV to 5 
GeV.  
The dependence of relic density with $(m_{N})_{11}$ and $M^{H}_{N}$ is shown 
in figure~\ref{fig:paramdepMeVmajorana}(c) where the contours depicts the 
value of relic density. The nature of the contours can be explained as 
follows. An increase in $(m_{N})_{11}$ raises $M_{\snu^{i}_{DM}}$, and the 
relic density, too, increases linearly. On the other hand, the freeze-in 
relic density increases with $M^{H}_{N}$, since $(Y_{\nu})_{ij}$
(for i,j=2,3) and hence $\Theta_{\snu\,1j}$ also increase 
with $M^{H}_{N}$. Thus as we increase either $(m_{N})_{11}$ or $M^{H}_{N}$ 
the relic density increases.
One should notice that even when
a fine cancellation between $\mu\cot\beta$ and $A_{\nu}$ takes place
({\it{viz.}} in figure~\ref{fig:paramdepMeVmajorana}(c) 
$|A_{\nu}-\mu\cot\beta|\,\simeq\,0.38\,{\rm GeV}$), $M^{H}_{N}$ is 
allowed to be only up to a few $\textit{tens of GeV}$, depending on the value 
of $(m_{N})_{11}$. This is because, although $|A_{\nu}-\mu\cot\beta|$ 
is fixed at $0.38\,{\rm GeV}$, $(Y_{\nu})_{ij}$(for i,j=2,3) 
increases with an increasing $M^{H}_{N}$ and consequently 
$(A_{\nu}-\mu\cot\beta\,\mp\,M^{H}_{N})(Y_{\nu})_{ij}$ becomes 
$\geq\,10^{-8}\,{\rm GeV}$ for some value of $M^H_N$, thereby overclosing 
the universe. 

A rather interesting signature of Dirac sneutrino LSP comes in the form of 
long-lived $\tilde{\tau}_{1}$ (lightest $\tilde{\tau}$ mass eigenstate, which 
is dominantly right-handed) NLSPs~\cite{Banerjee:2016uyt, Abdallah:2018gjj, 
Ghosh:2017vhe, Banerjee:2018uut}. A $\tilde{\tau}_{1}$ NLSP decays to 
$\snu_{DM}$ in association with a $W^{\pm}$ boson. The decay width is given 
by,
\begin{equation}
\Gamma_{\tilde{\tau_{1}}}\,=\,\frac{g^2\Theta_{\snu\,13}^2}{32\,\pi}|U_{L1}^{(\tilde{\tau_{1}})}|^2\,\frac{M_{\tilde{\tau_{1}}}^3}{m_{W}^2}\,\left[1-\frac{2(M_{\snu_{DM}}^2+m_{W}^2)}{M_{\tilde{\tau_{1}}}^2}\,+\,\frac{(M_{\snu_{DM}}^2 - m_{W}^2)^2}{M_{\tilde{\tau_{1}}}^4}\,\right]^{3/2},
\label{eqn:staudecaywidth} 
\end{equation}
where $M_{\tilde{\tau}_{1}}$ is the $\tilde{\tau}_{1}$ mass, $m_{W}$ is the 
$W$-boson mass and $U^{\tilde{\tau_{1}}}_{L1}\,=\,\sin\theta_{\tilde{\tau}
_{1}}$. Where, the stau mixing angle $\theta_{\tilde{\tau}_{1}}$ is given 
by,
\begin{equation}
\tan\,2\theta_{\tilde{\tau}_{1}}\,=\,\frac{2\,y_{\tau}\,v\,\sin\beta\,|A_{\tau}-\mu\cot\beta|}{m^{2}_{\tilde{l}_{L}}-m_{\tilde{\tau}_{R}}}.
\label{eqn:staumixangle} 
\end{equation}
All the parameters given in equation~\eqref{eqn:staumixangle} are 
self-explanatory and their values can be found in figure~\ref{fig:paramdepMeVmajorana}(d). One may wonder whether a 
$\tilde{\tau}_{1}$ can be the NLSP in presence of $\Delta\,L=2$ masses, 
since the sneutrino mixing angles $\Theta_{\snu\,1j}$, which determines the 
lifetime of $\tilde{\tau}_{1}$, is severely constrained in the present 
scenario. Figure~\ref{fig:paramdepMeVmajorana}(d) shows the variation of
$\tilde{\tau}_{1}$ lifetime with its mass when right-sneutrino has correct 
relic density $\textit{i.e.}$ $\Theta_{\snu\,1j}\,\lesssim\,10^{-12}$.
Clearly a wide range of values of $M_{\tilde{\tau}_{1}}$ is allowed
where the lifetime of $\tilde{\tau}_{1}$ is $\leq\,100\,sec$, a constraint 
imposed by BBN. Thus even in presence of $\Delta\,L\,=2$ masses, right-
sneutrino can serve as a non-thermal DM candidate along with a 
$\tilde{\tau}_{1}$ NLSP, that leaves its footprints in the form of a 
$\textit{heavy stable charged track}$ inside collider detectors.

\subsection{Results for $m_{susy}\,\simeq\,\mathcal{O}(100)\,$TeV}
\label{sec:highsusy}
 
\begin{figure}[t]
\begin{center}
  \includegraphics[width=5cm,angle=270]{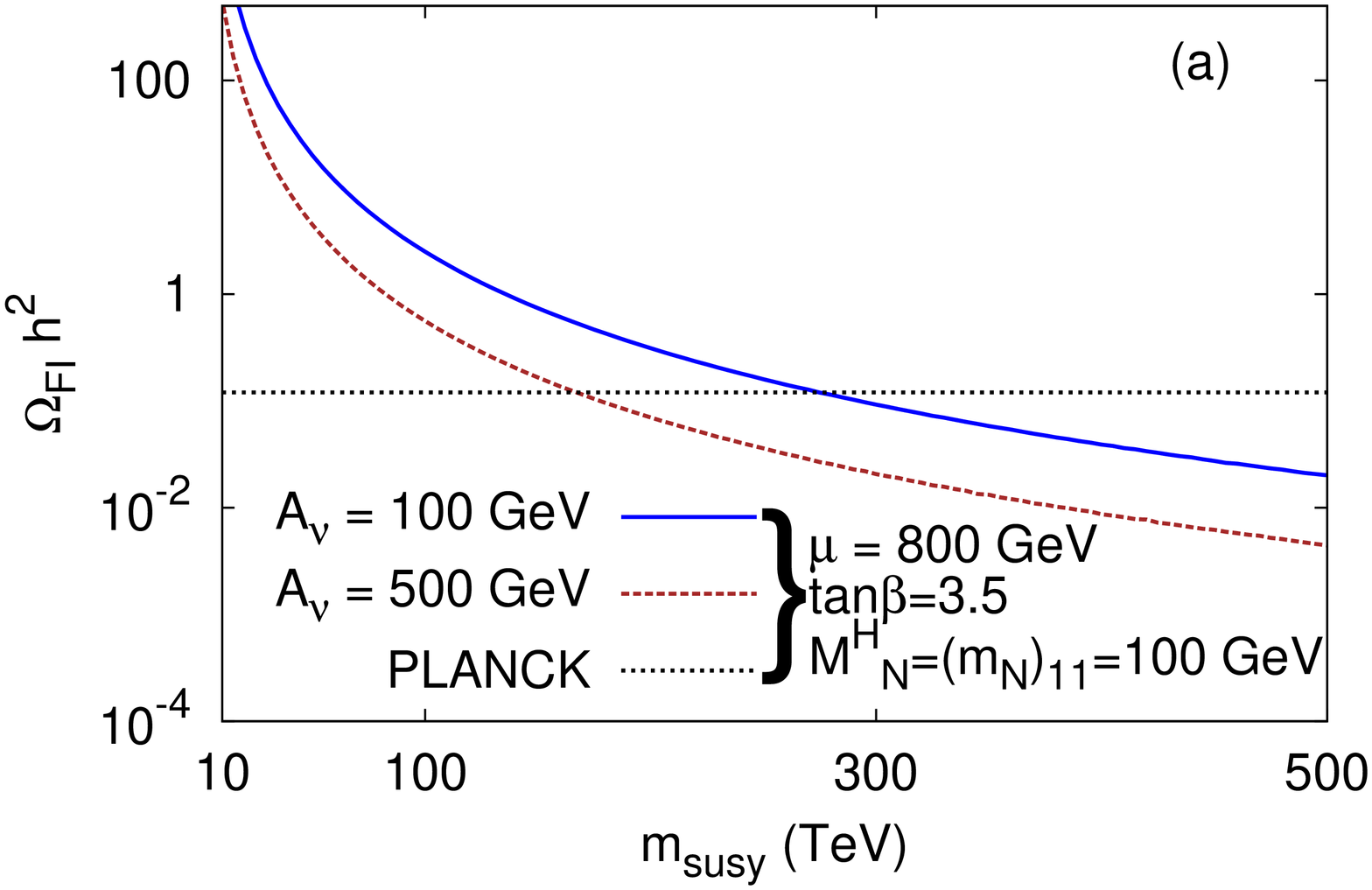}
  \includegraphics[width=5cm,angle=270]{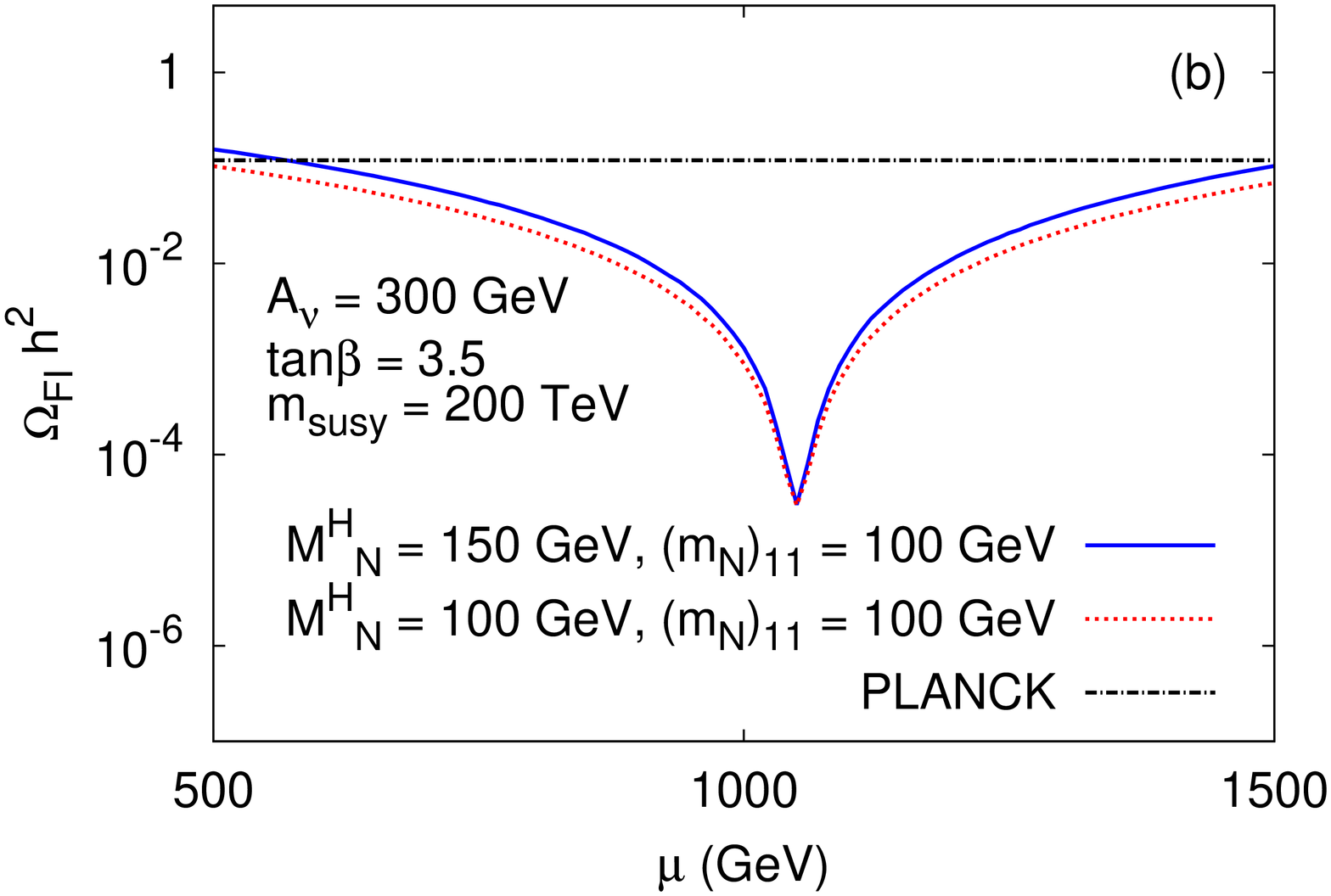}
  \includegraphics[width=5cm,angle=270]{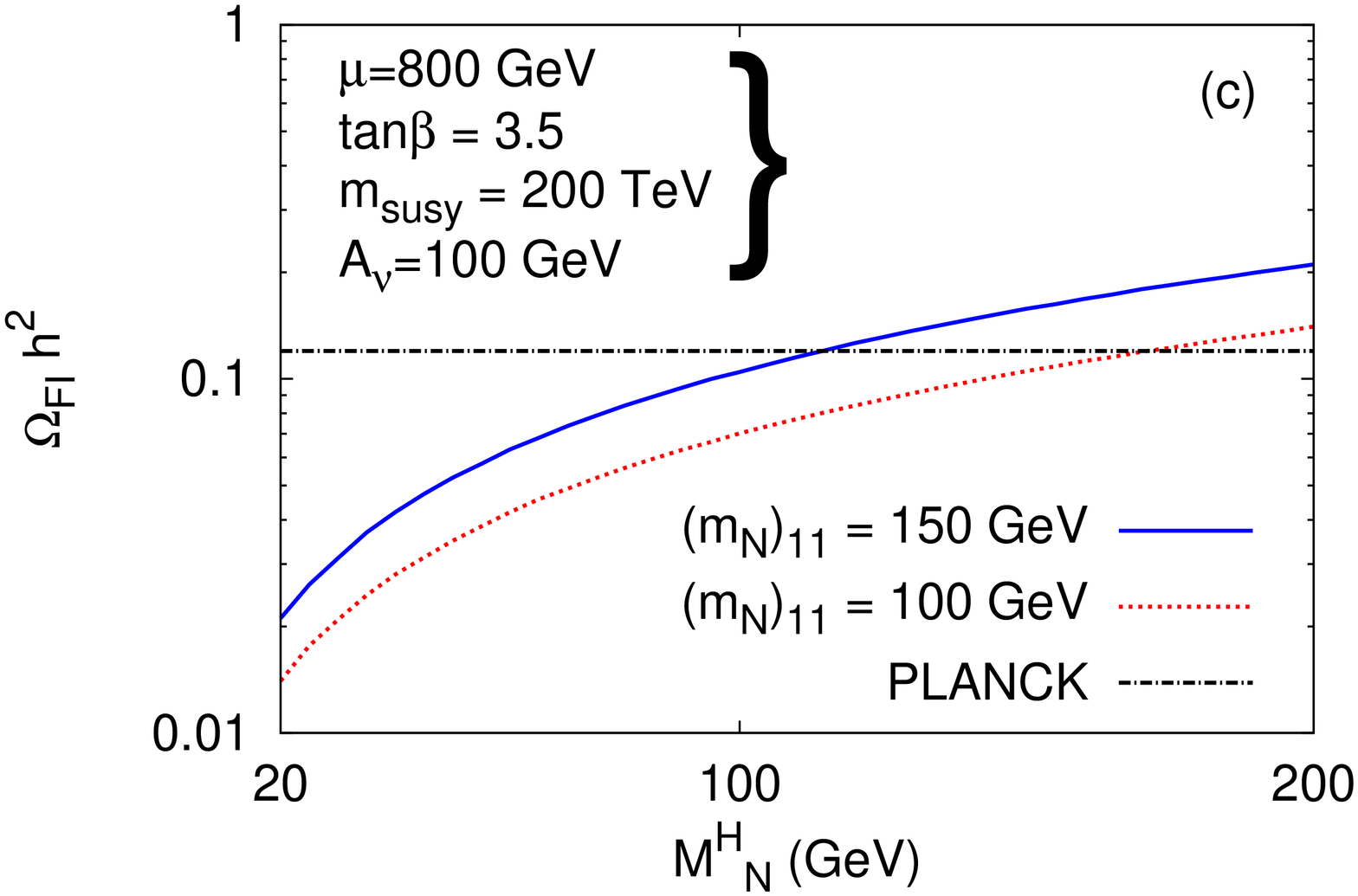}
  \includegraphics[width=5cm,angle=270]{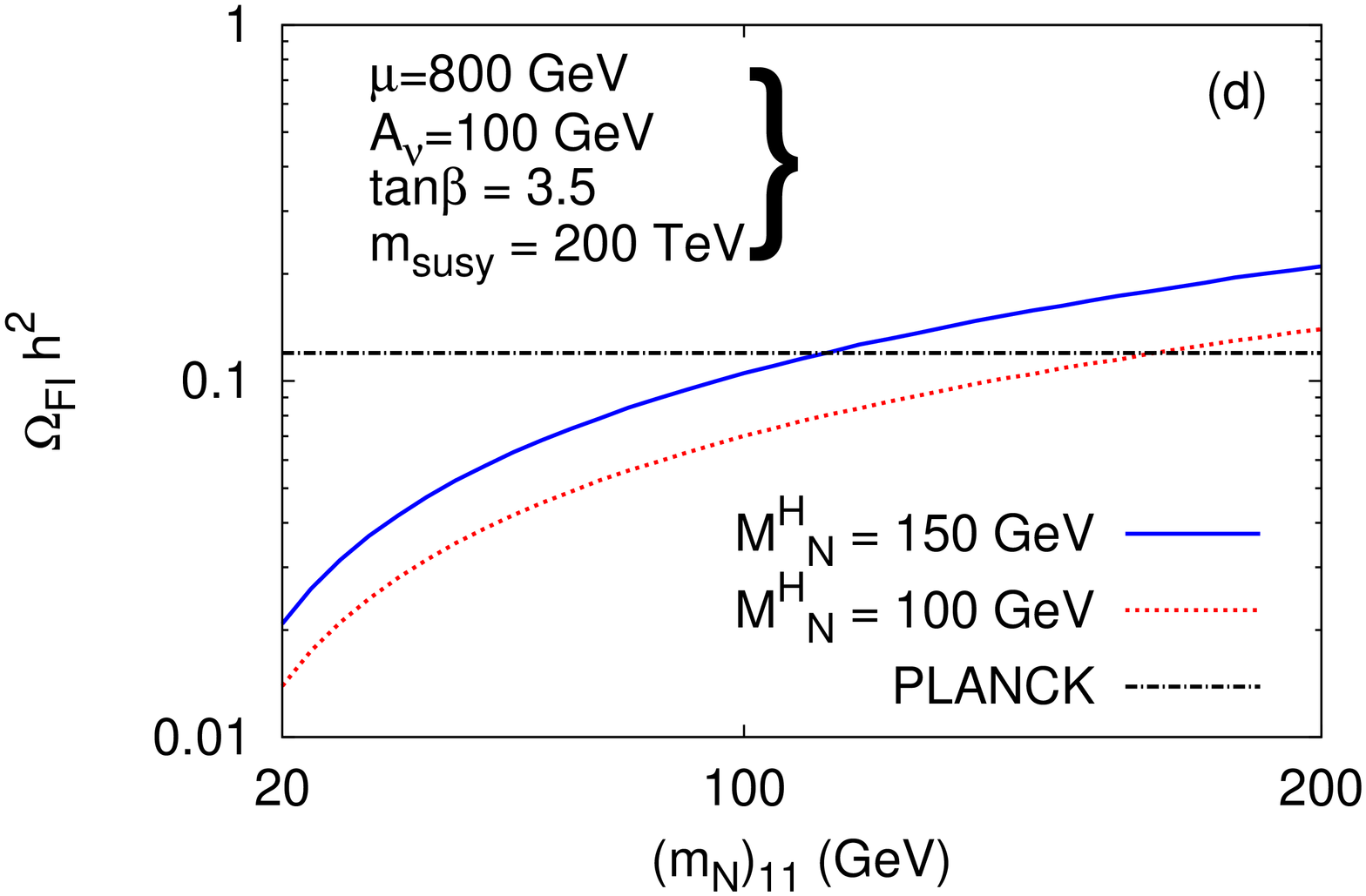}
\caption{Subfigure~\ref{fig:paramdepGeVmajorana}(a) show the variation of 
relic density a with increase in $m_{susy}$ for $A_{\nu}$,$\,\mu\,$,$
(m_{N})_{11}$ and $M^{H}_{N}$ around electroweak scale. 
In order to generate correct SM like higgs mass and satisfy all the 
constraints of EWSB $\tan\beta\,=3.5$ have been taken. 
Subfigure~\ref{fig:paramdepGeVmajorana}(b) depicts how the allowed region of 
$\mu-A_{\nu}$ plane have been extended due to such increase in SUSY-breaking 
masses $\textit{i.e.}\,m_{susy}$. The subfigure~\ref{fig:paramdepGeVmajorana}(c) and \ref{fig:paramdepGeVmajorana}(d)
plots of the lower panel show the variation of $\textit{freeze-in}$ relic 
density with $(m_{N})_{11}$ and $M^{H}_{N}$ respectively.}
\label{fig:paramdepGeVmajorana}
\end{center}
\end{figure} 

As we have seen, for $m_{susy}\simeq\,1\,{\rm TeV}$, Majorana masses up to 
a ${\rm \textit{few tens of GeV}}$ are allowed. 
If we allow even larger values for heavy neutrino Majorana masses 
($M^{H}_{N}$) then the Yukawa couplings also increase 
leading to an increase in the mixing angles $\Theta_{\snu\,1j}$. 
To compensate the effect we have to increase $m_{susy}$, and the 
denominator of the matrix $\Theta_{\snu}$ consequently increases. 
As an example, for superparticle masses around 100 TeV (or beyond), 
the relic density requirement can be expressed as 
\begin{eqnarray}
\underset{i}{\sum}\,M_{\snu^{i}_{DM}}\Gamma_{\snu^{i}_{DM}}\,\lesssim\,5.2\times 10^{-15}\,{\rm GeV^{2}}.
\label{eqn:highmsusyFIrelicbound}
\end{eqnarray}
If one assumes the DM masses to be still in the electroweak scale, the above 
equation implies $\Theta_{\snu\,1j}\,\simeq\,10^{-9}$.  
From equation~\eqref{eqn:rotang} it is evident that 
$M_{N}^H\,\simeq\,100\,$GeV will be allowed in this scenario. 

The figure~\ref{fig:paramdepGeVmajorana}(a) shows the 
dependence of relic density on $m_{susy}$ for $M^{H}_{N}\,=(m_{N})_{11}=\,100\,{\rm GeV}$. 
Clearly, for $m_{susy}\,\simeq\,100\,{\rm TeV}$, the relic density is 
satisfied depending on the value of $|A_{\nu}-\mu\cot\beta|$. 
Figure~\ref{fig:paramdepGeVmajorana}(b)depicts the dependence of relic 
density on $\mu$ (for a fixed $A_{\nu}$ and $\tan\beta$), which can differ 
from $A_{\nu}\tan\beta$ up to a few hundreds of GeV in this case. The 
dependency of relic density on $M^{H}_{N}$ and $(m_{N})_{11}$ 
is shown in Figure~\ref{fig:paramdepGeVmajorana}(c) and (d) respectively. 
Both the plots show the expected increase in relic density as 
we go on increasing $M^{H}_{N}$ and $(m_{N})_{11}$. The explanation for
the predicted behaviours can be found in the discussion of 
figure~\ref{fig:paramdepMeVmajorana} in the previous subsection.

One should note that $\mu$, $M_{\snu^{i}_{DM}}$ and $A_{\nu}$ have been 
varied around the electroweak scale with $\tan\beta\,=\,3.5$.This is because
electroweak symmetry breaking conditions are difficult to satisfy with larger 
$\tan\beta$, if $m_{susy}$ is as high as 100 TeV. It is quite understandable
that $M_{\snu^{i}_{DM}}\,$ should be of $\mathcal{O}(100\,{\rm GeV})$ in 
order to obtain correct relic density (following figure~\ref{fig:paramdepGeVmajorana}(d)). 
On the other hand, $\mu$ should also 
be within a TeV, not only for satisfying electroweak symmetry breaking 
conditions, but also to control freeze-in via $\snu_{L}\rightarrow\snu_{R}\,h
$. And consequently a $A_{\nu}\,\simeq\,100\,{\rm GeV}$ is also required
so that $|A_{\nu}-\mu\cot\beta|$ does not become large enough to make the 
universe overclose. Hence, in this scenario we face an inexplicable hierarchy 
of SUSY-breaking parameters, to obtain the correct relic density, unless one 
engineer some highly contrived fine-tuning between $A_{\nu}$ and 
$\mu\cot\beta$ (see subsection~\ref{sec:MeVGeVmajo}).
    
\section{Summary and conclusions}\label{sec:summary}
We have worked with an MSSM scenario augmented with three families of 
right-chiral neutrino superfields, and $\Delta\,L=2$ mass terms in the 
superpotential as well as the scalar potential where masses for the light 
neutrinos have been generated following Type-I seesaw mechanism.
The right-handed sneutrinos are non-thermal dark matter candidate as a result 
of small Yukawa couplings. We have restricted to hierarchical neutrino 
masses, latest constraints from Planck data on DM as well as data on the 
neutrino sector. We have gone beyond earlier studies~\cite{Gopalakrishna:2006kr}
where constraints on the neutrino Majorana masses were obtained, mostly 
within a degenerate neutrino scenario. While the eV-scale scenario proposed 
there is presently disfavoured, we have shown that the current picture admits 
of considerable varieties as well as constraints in view of current 
observations. 
\begin{itemize}
\item While the lightest neutrino mass and consequently the corresponding 
entry in the (diagonal) right-handed neutrino mass matrix is a free 
parameter, the remaining $\Delta\,L=\,2$ mass terms are constrained rather 
tightly.
\item If all three $\Delta\,L=\,2$ entries in $M_{N}$ are in the keV range, 
there is an allowed region of parameter space to realize a
right-handed sneutrino as non-thermal DM. But the all three sterile 
neutrinos, too, effectively constitute warm dark matter. 
Such a situation, however, is disfavoured by the Dodelson-Widrow mechanism 
where the abundance of the warm DM goes up unacceptably.
\item For $\Delta\,L=\,2$ masses in the range 500 MeV-a few GeV, all 
constraints can be satisfied only if $A_{\nu}$ and the $\mu$-parameter are 
considerably fine-tuned.
\item  $\Delta\,L=\,2$ masses on the electroweak scale and above are possible 
only if DM mass $M_{\snu^{i}_{DM}}$, $\mu$ and $A_{\nu}$ are 2-3 orders of 
magnitude smaller than the remaining SUSY-breaking parameters, the later 
being on the order of 100 TeV.
\end{itemize}
We thus conclude that the right-sneutrino DM scenario, if at all the picture 
of nature, is subject to rather severe constraints in presence of
the $\Delta\,L=\,2$ masses. The exception to this requires an inexplicable 
hierarchy among SUSY-breaking mass parameters or a severe cancellation 
between SUSY-conserving $\mu$ and SUSY-breaking parameter $A_{\nu}$. 
\section{Acknowledgements}\label{sec:Acknowledgements}
We thank Raj Gandhi and Saptarshi Roy for for helpful discussions.
AG and BM acknowledges the hospitality of the Theoretical
Physics Department, Indian Association for Cultivation of Science, Kolkata, 
where the last part of the project was carried out. This work was partially 
supported by funding available from the Department of Atomic Energy, 
Government of India, for the Regional Centre for Accelerator-based Particle 
Physics (RECAPP), Harish-Chandra Research Institute.
\appendix 
\section{Decay Widths}\label{sec:Appendix1}
The decay widths of the channels that contribute to \textit{freeze-in} of 
$\snu_{DM}$ are as follows~\cite{Baer:2006rs}: 
\begin{subequations}
\begin{align}
\Gamma\left({\tilde{H}^{0}_{u}\,\rightarrow\,\snu_{R}\nu_{L}}\right) & =  \frac{1}{32\pi\,M_{\tilde{H}^{0}_{u}}}\beta_{f}(M_{\tilde{H}^{0}_{u}},M_{\nu_{L}})(M^{2}_{\tilde{H}^{0}_{u}}+M^{2}_{\nu_{L}}-M^{2}_{\snu_{R}})|\mathcal{A}\left({\tilde{H}^{0}_{u}\,\rightarrow\,\snu_{R}\nu_{L}}\right)|^{2},\\
\Gamma\left({\tilde{H}^{+}_{u}\,\rightarrow\,\snu_{R}l^{+}}\right) & =  \frac{1}{32\pi\,M_{\tilde{H}^{+}_{u}}}\beta_{f}(M_{\tilde{H}^{+}},M_{l})(M^{2}_{\tilde{H}^{+}_{u}}+M^{2}_{l}-M^{2}_{\snu_{R}})|\mathcal{A}\left({\tilde{H}^{+}_{u}\,\rightarrow\,\snu_{R}l^{+}}\right)|^{2},\\
\Gamma\left({\snu_{L}\,\rightarrow\,\snu_{R}h}\right) & =  \frac{1}{32\pi\,M_{\snu_{L}}}\beta_{f}(M_{\snu_{L}},M_{h})|\mathcal{A}\left({\snu_{L}\,\rightarrow\,\snu_{R}h}\right)|^{2},\\
\Gamma\left({\snu_{L}\,\rightarrow\,\snu_{R}Z}\right) & =  \frac{1}{32\pi\,M_{\snu_{L}}}\frac{M^{4}_{\snu_{L}}}{M^{2}_{Z}}\beta^{3}_{f}(M_{\snu_{L}},M_{Z})|\mathcal{A}\left({\snu_{L}\,\rightarrow\,\snu_{R}Z}\right)|^{2},\\
\Gamma\left({\tilde{l}_{L}\,\rightarrow\,\snu_{R}W^{-}}\right) & =  \frac{1}{32\pi\,M_{\tilde{l}_{L}}}\frac{M^{4}_{\tilde{l}_{L}}}{M^{2}_{W}}\beta^{3}_{f}(M_{\tilde{l}_{L}},M_{W})|\mathcal{A}\left({\tilde{l}_{L}\,\rightarrow\,\snu_{R}W^{-}}\right)|^{2},
\label{eqn:decays1}\\
\Gamma\left({\tilde{B}^{0}\,\rightarrow\,\snu_{R}\nu_{L}}\right) & =  \frac{1}{32\pi\,M_{\tilde{B}^{0}}}\beta_{f}(M_{\tilde{B}^{0}},M_{\nu_{L}})(M^{2}_{\tilde{B}^{0}}+M^{2}_{\nu_{L}}-M^{2}_{\snu_{R}})|\mathcal{A}\left({\tilde{B}^{0}\,\rightarrow\,\snu_{R}\nu_{L}}\right)|^{2},
\label{eqn:decays2}\\
\Gamma\left({\tilde{W}^{0}\,\rightarrow\,\snu_{R}\nu_{L}}\right) & =  \frac{1}{32\pi\,M_{\tilde{W}^{0}}}\beta_{f}(M_{\tilde{W}^{0}},M_{\nu_{L}})(M^{2}_{\tilde{W}^{0}}+M^{2}_{\nu_{L}}-M^{2}_{\snu_{R}})|\mathcal{A}\left({\tilde{W}^{0}\,\rightarrow\,\snu_{R}\nu_{L}}\right)|^{2},\\
\Gamma\left({\tilde{W}^{+}\,\rightarrow\,\snu_{R}l^{+}}\right) & =  \frac{1}{32\pi\,M_{\tilde{W}^{+}}}\beta_{f}(M_{\tilde{W}^{+}},M_{l})(M^{2}_{\tilde{W}^{+}}+M^{2}_{l}-M^{2}_{\snu_{R}})|\mathcal{A}\left({\tilde{W}^{+}\,\rightarrow\,\snu_{R}l^{+}}\right)|^{2}.
\end{align}
\end{subequations}
where $\beta_{f}(M_{A},M_{B})\,=\,\left(1-2\frac{M^{2}_{\snu_{R}}+M^{2}
_{B}}{M^{2}_{A}} + \frac{(M^{2}_{\snu_{R}}\,-\,M^{2}_{B})^{2}}{M^{4}_{A}}
\right)^{1/2}$ and $\mathcal{A}_{A\,\rightarrow\,B\,\snu_{R}}$ is the 
amplitude for the decay $A\rightarrow\,B\,\snu_{R}$. We have tabulated 
all the relevant amplitudes in ~\ref{sec:Appendix2}.

\section{Amplitudes}\label{sec:Appendix2}
In this section we give the expressions for the amplitudes for all the 
decays considered for calculation of $\textit{freeze-in}$ relic.
\begin{enumerate}
\item The two blocks of the sneutrino mass-matrix $M_{\snu}$ given in
equation~\eqref{realsnu2} are diagonalized by two different hermitian
matrices. We denote both of these matrices as $Z_{\snu}$ and use the
appropriate matrix depending on whether we are calculating the 
relic density of CP-even or CP-odd $\snu_{R}$ LSP. It is quite clear
that there being 6 CP-even and 6 CP-odd sneutrinos each of these $Z_{\snu}$
are $6\times6$ matrices which one can cast in terms of the sneutrino 
mixing matrix $\Theta_{\snu}$ as,
\begin{equation}
Z_{\snu}\,\approx\,
\begin{bmatrix}
\mathbb{I} & -\Theta^{T}_{\snu} \\
\Theta_{\snu} & \mathbb{I} \\
\end{bmatrix}
\end{equation}
where $\mathbb{I}$ is $3\times3$ identity matrix. The mixing of the 
lightest sneutrino eigenstate with the left-handed flavour eigenstate 
sneutrinos are given by $Z_{\snu\,4k}\,=\,\Theta_{\snu\,1k}$ with 
k=1,2,3 and the corresponding mixing with right-chiral sneutrinos are 
given by $Z_{\snu\,4(3+k)}\,=\,\delta_{1k}$ where k=1,2,3.
\item The full $6\times6$ mass-matrices of neutrinos ($\nu_{Li},N_{Ri}$ with 
i=1,2,3) is given by, 
\begin{equation}
M_{\nu}=
\begin{bmatrix}
0 & m^{T}_{D} \\
m_{D} & M_{N} \\
\end{bmatrix}
\end{equation}
which is diagonalized by a $6\times6$ matrix,
\begin{equation}
U_{V}=
\begin{bmatrix}
U_{0} &  S \\
S  &  U_{H} \\
\end{bmatrix}
\end{equation}
where the top $3\times3$ block denoted by $U_{0}$ is the usual PMNS-matrix.
$S$ denotes the mixing between $\nu_{L}$s and $N_{R}$s and $U_{H}$ denotes
the mixing between different generations of $N_{R}$s. 
\end{enumerate}
In the expressions that follows among the free indices $i\,,j$ 
denotes the sneutrino generation appearing in the vertex while 
$I$ denotes the same for R-parity even particles and sleptons. 
Thus $i,j=1,...,6$ and $I=1,2,3$ are allowed. 
Where index $i\,=\,4$ corresponds to sneutrino LSP. 
The indices which has to be summed over have been shown explicitely. 
$P_{L}\,=\dfrac{1-\gamma_{5}}{2}$ is the left-handed projection operator, 
$g_{2}$ being the weak coupling of SM and $\theta_{w}$ is the Wienberg 
angle.
\subsection{Interactions with Gauginos}
\begin{enumerate}
\item $\mathcal{A}\left({\tilde{B}^{0}\,\rightarrow\,\snu^{i}_{R}\nu^{I}_{L}}\right)\,
=\,\dfrac{i}{2}\,g_{1}\,P_{L}\,\underset{k=1}{\overset{3}{\sum}}\,U^{*}
_{0\,Ik}\,Z_{\snu\,ik}$
\item $\mathcal{A}\left({\tilde{W}^{0}\,\rightarrow\,\snu^{i}_{R}\nu^{I}_{L}}\right)\,
=\,-\dfrac{i}{2}\,g_{2}\,P_{L}\,\underset{k=1}{\overset{3}{\sum}}\,U^{*}
_{0\,Ik}\,Z_{\snu\,ik}$
\item $\mathcal{A}\left({\tilde{H}^{0}_{u}\,\rightarrow\,\snu^{i}_{R}\nu^{I}_{L}}\right)\,
=\,-\dfrac{i}{\sqrt{2}}\,P_{L}\,\left(\,\underset{k,l=1}{\overset{3}
{\sum}}\,Z^{*}_{\snu\,ik}\,Y_{\nu\,lk}\,U_{V\,I(3+l)}\,+\,\underset{k,l=1}
{\overset{3}{\sum}}\,Z^{*}_{\snu\,i(3+l)}\,Y_{\nu\,lk}\,U_{V\,Ik}\,\right)$
\item $\mathcal{A}\left({\tilde{W}^{-}\,\rightarrow\,\snu^{i}_{R}l^{-\,I}}\right)\,
=\,-\dfrac{i}{2}\,g_{2}\,P_{L}\,Z^{*}_{\snu\,iI}$
\item $\mathcal{A}\left({\tilde{H}^{+}_{u}\,\rightarrow\,\snu^{i}_{R}l^{-\,I}}\right)\,
=\,\dfrac{i}{2}\,P_{L}\,\underset{k=1}{\overset{3}{\sum}}Z^{*}_{\snu
\,i(3+k)}\,Y_{\nu\,kI}$
\end{enumerate}

\subsection{Interactions with sfermions}
Sneutrino LSP can decay from sleptons and heavier sneutrinos via SM gauge 
bosons or SM higgs which we have discussed below: 
\subsubsection{Interactions with Gauge-Bosons}
\begin{enumerate}
\item $\mathcal{A}\left({\tilde{e}^{I}_{L}\,\rightarrow\,\snu^{i}_{R}W^{\mu}}\right)\,
=\,\dfrac{i}{2}\,g_{2}\,Z^{*}_{\snu\,iI}\,\left(\,p_{\snu^{i}}-p_{\tilde{e}
_{L}}\,\right)^{\mu}$
\item $\mathcal{A}\left({\snu_{H}^{j}\,\rightarrow\,\snu^{i}_{R}Z^{\mu}}\right)\,
=\,\dfrac{1}{2}\,\left(g_{2}\cos\theta_{w}+g_{1}\sin\theta_{w}\,\right)\,
\left(\,p_{\snu_{H}}-p_{\snu^{i}}\,\right)^{\mu}\,\underset{k=1}
{\overset{3}{\sum}}\,Z^{*\,H}_{\snu\,jk}\,Z^{*}_{\snu\,ik}$

here $H$ denotes the heavier sneutrino eigenstates and hence j runs from 
1 to 5. One has to keep in mind that this vertex is non-zero only when 
the CP nature of the two participating sneutrinos($\snu_{H}^{j}$ and 
$\snu^{i}$) are opposite. Hence if $Z_{\snu}$ is the diagonalizing 
matrix corresponding to CP-even states then $Z^H_{\snu}$ has to be that 
corresponding to CP-odd states. 
\end{enumerate}

\subsubsection{Interactions with SM higgs}
\begin{enumerate}
\item For a heavier CP-even(or CP-odd) sneutrino($\snu^{j}_{H}$) decaying 
into a lighter CP-even(or CP-odd) sneutrino ($\snu^{i}$),
\begin{eqnarray}
\mathcal{A}\left({\snu_{H}^{j}\,\rightarrow\,\snu^{i}_{R}h}\right)\,
&=&\,\dfrac{i}{2\sqrt{2}}\,\Bigg(\mu\,\underset{p,k=1}{\overset{3}{\sum}}
\left[Z^{*\,H}_{\snu\,jk}\,Z^{*}_{\snu\,i(3+p)}\,+\,Z^{*\,H}_{\snu\,j(3+p)}
\,Z^{*}_{\snu\,ik}\right]Y^{*}_{\nu\,pk}
  \, \nonumber \\
  && \hspace{0.7cm}\, + \mu^{*}\,\underset{p,k=1}{\overset{3}{\sum}}
\left[Z^{*\,H}_{\snu\,jk}\,Z^{*}_{\snu\,i(3+p)}\,+\,Z^{*\,H}_{\snu\,j(3+p)}
\,Z^{*}_{\snu\,ik}\right]\,Y_{\nu\,pk}\,\Bigg)\, \nonumber \\
&& -\,2(g^{2}_{1}+g^{2}_{1})\,v_{d}\,\underset{k=1}{\overset{3}{\sum}}\,Z^{*
\,H}_{\snu\,jk}\,Z^{*}_{\snu\,ik} \nonumber
\end{eqnarray}

For higgs-mediated sneutrino decays also j runs from 1 to 5  as in the case 
of Z-mediated processes and in this case both $Z_{\snu}$ and 
$Z^H_{\snu}$ have to be the mixing matrix corresponding to CP-even
(or CP-odd) states. 

In diagonalizing sneutrino mass matrix and calculating all the Feynman 
Rules we have used $\texttt{SARAH}$~\cite{Staub:2013tta}. 
\end{enumerate}


\bibliographystyle{apsrev4-1}
\bibliography{right_snu}

\end{document}